\documentclass[traditabstract]{aa}

\usepackage{graphicx}
\usepackage{hyperref}
\usepackage{txfonts}

\usepackage{amsmath}
\usepackage{amssymb}
\usepackage{graphicx}
\usepackage{newtxtext,newtxmath}
\usepackage{Times}

\usepackage{ae,aecompl}
\usepackage[T1]{fontenc}

\usepackage{bm}
\usepackage{booktabs}
\usepackage{lipsum}
\usepackage{multirow}
\usepackage{natbib}
\usepackage{soul}

\newcommand{\add}[1]{{#1}}
\newcommand{\citeprep}[1]{{\color{blue}#1}}

\newcommand{\EWoii}{$EW$[O{\sc ii}] }
\newcommand{\msun}[1]{$10^{#1}\mathrm{M_\odot}$}
\newcommand{\viscos}{VIS\textsuperscript{3}COS}
\makeatletter
\newcommand*{\citelink}[1]{\hyper@link{cite}{cite.#1}}
\makeatother
\newcommand{\citeAPA}{\citelink{paulino-afonso2018a}{PA18}}

\bibpunct[ ]{(}{)}{;}{a}{}{,}

\hypersetup{
     colorlinks   = True,
     citecolor    = blue,
     linkcolor = blue,
     urlcolor = blue
}

\begin{document} 


\title{VIS$\boldsymbol{^3}$COS: III. environmental effects on the star formation histories of galaxies at $z\sim0.8$ seen in [O{\sc ii}], H$\delta$, and $D_n4000$\thanks{Based on observations obtained with VIMOS on the ESO/VLT under the programmes 086.A-0895, 088.A-0550, and 090.A-0401.}}

\author{
Ana Paulino-Afonso \inst{1,2,3}\fnmsep\thanks{E-mail: aafonso@oal.ul.pt} \and
David Sobral \inst{3}
\and
Behnam Darvish \inst{4} 
\and 
Bruno Ribeiro \inst{5}
\and
Ian Smail \inst{6}
\and
Philip Best\inst{7}
\and
Andra Stroe \inst{8} 
\and 
Joseph Cairns \inst{3}
}

\institute{Instituto de Astrof\'isica e Ci\^encias do Espa\c{c}o, Universidade de Lisboa, OAL, Tapada da Ajuda, PT1349-018 Lisboa, Portugal
\and
Departamento de F\'isica, Faculdade de Ci\^encias, Universidade de Lisboa, Edif\'icio C8, Campo Grande, PT1749-016 Lisboa, Portugal
\and
Department of Physics, Lancaster University, Lancaster, LA1 4YB, UK
\and
Cahill Center for Astrophysics, California Institute of Technology, 1216 East California Boulevard, Passadena, CA 91125, USA
\and
Leiden Observatory, Leiden University, P.O. Box 9513, NL-2300 RA Leiden, The Netherlands
\and
Centre for Extragalactic Astrophysics, Department of Physics, Durham University, Durham DH1 3LE, UK
\and
Institute for Astronomy, University of Edinburgh, Royal Observatory, Blackford Hill, Edinburgh EH9 3HJ, UK
\and
Harvard-Smithsonian Center for Astrophysics, 60 Garden Street, Cambridge, MA 02138, USA
}

\date{ } 

\abstract{We present spectroscopic observations of 466 galaxies in and around a superstructure at $z\sim0.84$ targeted by the VIMOS Spectroscopic Survey of a Supercluster in the COSMOS field (VIS$^{3}$COS). We use [O{\sc ii}]$\lambda$3727, H$\delta$, and $D_n4000$ to trace the recent, mid-, and long-term star formation histories and investigate how stellar mass and the local environment impacts those. By studying trends both in individual and composite galaxy spectra, we find that both stellar mass and environment play a role in the observed galactic properties. Low stellar mass galaxies ($10<\log_{10}\left(M_\star/\mathrm{M_\odot}\right)<10.5$) in the field show the strongest H$\delta$ absorption. Similarly, the massive population ($\log_{10}\left(M_\star/\mathrm{M_\odot}\right)>11$) shows an increase in H$\delta$ absorption strengths in intermediate-density environments (e.g. filaments). Intermediate stellar mass galaxies ($10.5<\log_{10}\left(M_\star/\mathrm{M_\odot}\right)<11$) have similar H$\delta$ absorption profiles in all environments, but show a hint of enhanced [O{\sc ii}] emission at intermediate-density environments. This hints that low stellar mass field galaxies and high stellar mass filament galaxies are more likely to have experienced a recent burst of star formation, while galaxies of the intermediate stellar-mass show an increase of star formation at filament-like densities. We also find that the median [O{\sc ii}] equivalent width (|EW$_\mathrm{[OII]}|$) decreases from $27\pm2$ \AA\ to $2.0_{-0.4}^{+0.5}$ \AA\ and $D_n4000$ increases from $1.09\pm0.01$ to $1.56\pm0.03$ with increasing stellar mass (from $\sim10^{9.25}$ to $\sim10^{11.35}\ \mathrm{M_\odot}$). Concerning the dependence on the environment, we find that at fixed stellar mass |EW$_\mathrm{[OII]}|$ is tentatively lower in higher density environments. Regarding $D_n4000$, we find that the increase with stellar mass is sharper in denser environments, hinting that such environments may accelerate galaxy evolution. Moreover, we find larger $D_n4000$ values in denser environments at fixed stellar mass, suggesting that galaxies are on average older and/or more metal-rich in such dense environments.
This set of tracers depicts a scenario where the most massive galaxies have, on average, the lowest sSFRs and the oldest stellar populations (age $\gtrsim1$ Gyr, showing a mass-downsizing effect). We also hypothesize that the observed increase in star formation (higher EW$_\mathrm{[OII]|}$, higher sSFR) at intermediate densities may lead to quenching since we find the quenched fraction to increase sharply from the filament to cluster-like regions at similar stellar masses.}


\authorrunning{A. Paulino-Afonso et al.}
\titlerunning{Stellar mass and environmental effects on [O{\sc ii}], H$\delta$, and $D_n4000$ at $z\sim0.8$}

\maketitle



\section{Introduction}\label{section:introduction}

One of the key tracers of galactic evolution is the rate at which gas is converted into stars, measured as the star formation rate \citep[SFR, e.g.][]{kennicutt1998,kennicutt2012}. Observations show that typical galaxies were actively forming stars at a rate of $\sim10$ times higher at $z\sim2$ than at $z\sim0$ \citep[both the cosmic star formation rate density and typical SFRs decrease during this epoch, see e.g.][]{madau2014,sobral2014}. One of the fundamental questions of modern Astronomy is to understand the mechanisms responsible for the regulation of star formation in galaxies and find how efficient galaxies are in converting gas into stars \citep[see e.g.][]{combes2013,lehnert2013}.

There are two broad groups of processes, internal and external, that can contribute to the evolution of any given galaxy \citep[e.g.][]{kormendy2013}. However, the contribution of each set of processes to regulating star formation in galaxies is still unclear \citep[see e.g.][]{erfanianfar2016}. Internal processes include dynamical instabilities \citep[e.g.][]{kormendy2013}, halo quenching \citep[e.g.][]{birnboim2003,keres2005,keres2009,dekel2006}, supernova feedback \citep[e.g.][]{efstathiou2000,cox2006}, and active galactic nuclei (AGN) feedback \citep[e.g.][]{bower2006,croton2006,somerville2008,fabian2012}. External processes include galaxy interactions with other galaxies or the inter galactic medium, specifically ram pressure stripping \citep[e.g.][]{gunn1972}, galaxy strangulation \citep[e.g.][]{larson1980,balogh2000}, galaxy-galaxy interactions and harassment \citep[e.g.][]{mihos1996,moore1998}, or tidal interaction between the large-scale gravitational potential and the galaxy \citep[e.g.][]{merritt1984,fujita1998}. These range of physical processes are thought of being the way through which galaxies regulate, and eventually halt the formation of new stars, a phenomenon commonly referred to as galaxy quenching \citep[e.g.][]{gabor2010,peng2010b}.

How each proposed mechanism affects individual galaxies is complex. In internal processes, feedback can either heat or eject the gas from galaxies preventing it from condensing in molecular clouds to form new stars \citep[e.g.][]{keres2009b}. Supernova feedback is thought to be more important at lower stellar masses, and AGN feedback is arguably an important mechanism for quenching at high stellar masses \citep[e.g.][]{puchwein2013}. Halo quenching refers to gravitational heating, preventing gas from cooling down and forming new stars. However, it requires a sustained mechanism to heat the gas \citep[e.g.][]{birnboim2007}, and cold gas flows might still penetrate the halo into the galactic disk to fuel star formation \citep[e.g.][]{keres2009}.

In terms of external processes, ram pressure stripping can initially compress the gas/dust thus increasing the column density of the gas and dust which is favourable for star formation \citep[e.g.][]{gallazzi2009,bekki2009,owers2012,roediger2014}. Tidal galaxy-galaxy interactions can lead to the compression and inflow of the gas in the periphery of galaxies into the central parts, feeding and rejuvenating the stellar populations in the central galactic regions which result in an enhancement in star formation activity \citep[e.g.][]{mihos1996,kewley2006,ellison2008}. Such encounters are probable when the galaxies have low relative velocities (i.e., low-velocity-dispersion environments) and are closer to each other (denser regions). Intermediate-density environments such as galaxy groups, in-falling regions of clusters, cluster outskirts, merging clusters, and galaxy filaments provide the ideal conditions for such interactions \citep[e.g.][]{moss2006, perez2009, li2009, sobral2011, tonnesen2012, darvish2014, stroe2014, stroe2015a, malavasi2017}. This enhancement of star formation is thought to be responsible for the subsequent quenching since most of the available gas is consumed or expelled through outflows in a short period of time effectively preventing future star formation from occurring in the galaxy without further external influence. If those events are ubiquitous, one should observe a temporary rise in star formation in such environments, after which galaxies are expected to become passive. This has been found in several studies, referring to the intermediate-density environments as sites of enhanced star formation rate and obscured star formation activity \citep[e.g.][]{smail1999, best2004, koyama2008, koyama2010, koyama2013, gallazzi2009, geach2009, sobral2011, sobral2016, coppin2012, stroe2015a, stroe2017}. 

Since many of these mechanisms are linked to the increased density of galaxies, it is natural to look in over-dense regions for the impact of the local environment on the observed properties of galaxy populations. In the local Universe ($z\sim0$), star formation is typically lower in higher density environments \citep[e.g.][]{oemler1974, dressler1980, lewis2002, kauffmann2004, blanton2005b, li2006, peng2010b, darvish2016, darvish2018}. By separating galaxies into distinct populations (star-forming and quiescent) studies find that the quenched fraction is highly dependent on the local density, at least up to $z\lesssim1$, with the quenched population being more common in high-density regions and a higher fraction of star-forming galaxies found in lower density regions \citep[e.g.][]{kodama2001,kodama2004,best2004,nantais2013,darvish2016,erfanianfar2016,cohen2017}. While in the local Universe the picture is clear, with the average star formation being the lowest in high-density, relaxed cluster regions \citep[e.g.][]{balogh2000,kauffmann2004}, it is still unclear if that holds at higher redshifts. Some studies find a flattening and/or reversal of such relation \citep[$z\sim1-1.5$, e.g.][]{cucciati2006,elbaz2007,ideue2009,tran2010,popesso2011,li2011,santos2014,stach2017,cooke2019}, while others find the same trends we see locally \citep[e.g.][]{patel2009,sobral2011,muzzin2012,santos2013,scoville2013,darvish2016}. It is possible that reconciling the different observed trends requires more detailed analysis on other possible underlying relations, especially with stellar mass \citep[e.g.][]{peng2010b,sobral2011,muzzin2012,darvish2016}, and also controlling for other properties such as AGN fraction and dust content. An alternative explanation might be the stochastic nature of the formation of dense environments which can explain the observed differences as a natural cosmic variance. 

Recent studies in the literature are using spectral indices [O{\sc ii}], H$\delta$, and $D_n4000$ to probe the stellar population of galaxies at intermediate redshifts ($0.5 \lesssim z \lesssim 1.2$) due to their availability in the observed optical frame. All these indicators, when combined, can be used to distinguish actively star-forming, (post-)starburst, and old/passive galaxies since they should occupy different regions of the possible parameter space \citep[e.g.][]{couch1987,balogh1999,poggianti1999,poggianti2009,fritz2014}. The [O{\sc ii}]$\lambda$3737 emission traces on-going star formation \citep[{timescales of $\sim$10 Myr}, e.g.][]{couch1987,poggianti1999,kennicutt1998,kewley2004,poggianti2006}, however, it depends also on the metallicity and can be a poor tracer for dusty galaxies \citep[e.g.][]{kewley2004, yan2006, kocevski2011}. By measuring the [O{\sc ii}] equivalent width we can also crudely trace the specific SFR (sSFR) which is found to anti-correlate with stellar mass \citep[e.g.][]{bridge2015,cava2015,darvish2015} with more massive star-forming galaxies having lower [O{\sc ii}] equivalent widths. Additionally, higher density environments are found to depress [O{\sc ii}] emission \citep[e.g.][]{balogh1999,darvish2015}.
The H$_\delta$ line (and other strong Balmer absorption lines) can be indicative of a post-starburst phase \citep[$\approx100-1000$ Myr after the burst, e.g.][]{couch1987, balogh1999, poggianti1999, poggianti2009, dressler2004, vergani2010, mansheim2017}, if a strong absorption (typical of A stars, where hydrogen absorption is the strongest) is observed and no tracers of on-going star formation are found \citep[][]{couch1987}. Recently, \citet{wu2018} found that H$\delta$ equivalent width correlates with stellar mass, with more massive galaxies having weaker H$\delta$ absorption lines, but have no study on the impact of environment \citep[see also e.g.][\! for a similar result on passive galaxies]{siudek2017}. Finally, a measure of the flux break at $4000$ \AA\ ($D4000$ and $D_n4000$, as defined by \citealt{bruzual1983} and \citealt{balogh1999}, respectively), traces the age of the galaxy and also the stellar metallicity (especially for older systems) to a lesser degree. This break is produced by a combination of metal absorption on the atmosphere of old and cool stars and the lack of flux from young and hot OB stars \citep[e.g.][]{poggianti1997, kauffmann2003b} and so it is sensitive to the average age of the stellar population. The 4000 \AA\ break is also found to be stronger for higher stellar mass galaxies \citep[e.g.][]{muzzin2012,vergani2008,hernan-caballero2013,siudek2017,wu2018}, hinting at their older stellar populations, in an average sense. In terms of local density, \citet{muzzin2012} found that galaxies in cluster environments have on average stronger breaks than their field counterparts at similar stellar masses, which they argue it can be explained by the different fractions of star-forming and quiescent galaxies in different environments. 

We aim to investigate the influence of environment on the star formation history of galaxies using a number of spectral indicators \citep[e.g.][]{balogh1999,poggianti1999,poggianti2009,dressler2004,vergani2010,mansheim2017b,wu2018}. Due to the spectral coverage of the VIMOS Spectroscopic Survey of a Superstructure in the COSMOS field \citep[VIS$^3$COS,][\!, hereafter \citeAPA]{paulino-afonso2018a} we estimate the current and past star formation activity of galaxies using a combination of [O{\sc ii}] (tracing on-going star formation in moderate to high star-forming galaxies, $\lesssim$10Myr), H$\delta$ (probing star formation on intermediate timescales - 50 Myr to $\sim$1 Gyr prior to observation), and $D_n4000$ (probing the star formation history on longer timescales). We investigate this using spectroscopic observations of $\sim500$ galaxies in and around a superstructure at $z\sim0.84$ in the COSMOS field \citep[][\!, \citeAPA]{sobral2011} by probing a wide range of environments and stellar masses with a single survey.

This paper is organized as follows: in Section \ref{section:data} we briefly explain the survey and give some details on the data used. Section \ref{section:methods} details the stacking methods and the spectroscopic measurements. In Section \ref{section:results} we present the results from individual and stacked spectral properties. We discuss our findings in Section \ref{section:discussion}. Section \ref{section:conclusions} where presents the conclusions of our study. We use AB magnitudes \citep{oke1983}, a Chabrier \citep{Chabrier2003} initial mass function (IMF), and assume a $\Lambda$CDM cosmology with H$_{0}$=70 km s$^{-1}$Mpc$^{-1}$, $\Omega_{M}$=0.3 and $\Omega_{\Lambda}$=0.7. The physical scale at the redshift of the superstructure ($z\sim0.84$) is 7.63 kpc/\arcsec.


\section{The VIS${^3}$COS survey}\label{section:data}
\begin{figure*}
\centering
\includegraphics[width=\linewidth]{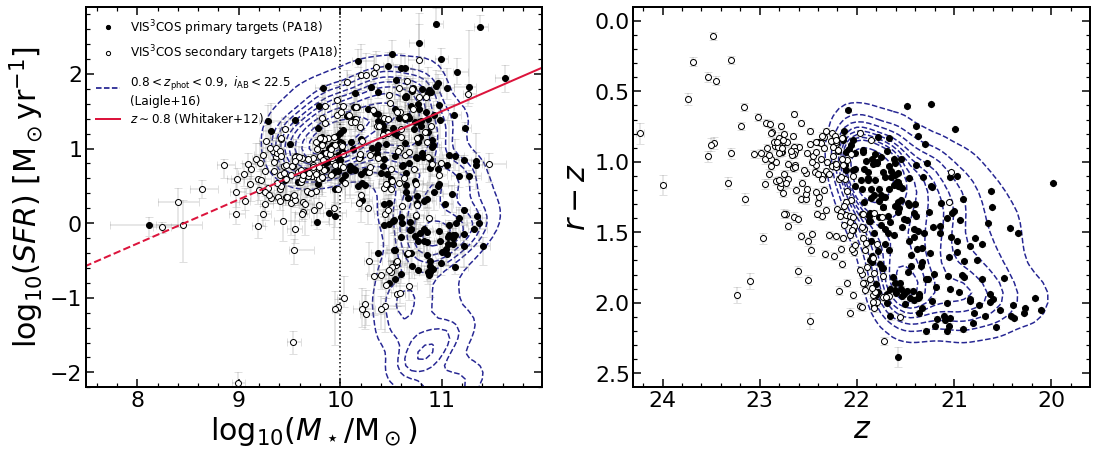}
\caption{Stellar masses and star formation rates derived from SED fitting in our spectroscopic sample at $0.8<z<0.9$, showing the primary (filled circles) and secondary (open circles) targets separately (left). Colour-magnitude diagram for the same sample (right). For comparison, we show the derived best-fit relation for star-forming galaxies computed at $z=0.84$ using the equation derived by \citet{whitaker2012} over a large average volume in the COSMOS field (the dashed line is an extrapolation below their stellar mass completeness). The vertical dotted line shows the approximate stellar-mass representativeness limit of our survey. The dotted contours show the COSMOS2015 distribution of galaxies with $0.8<z_\mathrm{phot}<0.9$ and $i_\mathrm{AB}<22.5$ from 10\% to 90\% of the sample in 10\% steps.}
\label{fig:mass_sfr_dist}
\end{figure*}

The VIS$\boldsymbol{^3}$COS survey maps a large $z\sim0.84$ over-density spanning 21\arcmin$\times$31\arcmin\ (9.6$\times$14.1 Mpc$^{2}$) in the COSMOS field  \citep{scoville2007} with the VIMOS instrument mounted on the VLT. This structure contains three confirmed X-ray clusters \citep{finoguenov2007} and also harbours a large-scale over-density of H$\alpha$ emitters \citep{sobral2011,darvish2014}. The full description of the data and redshift measurements are presented in \citeAPA , and we briefly describe here some details.

Our primary targets were selected from the \cite{ilbert2009} catalogue and with $0.8<z_\mathrm{phot,l}<0.9$ (with $z_\mathrm{phot,l}$ being either the upper or lower $99$\% confidence interval limit for each source) and $i_\mathrm{AB}<22.5$. To effectively fill the masks we also add as secondary targets galaxies down to $i_\mathrm{AB}<23$ and with photometric redshifts in the interval $0.6 < z < 1.1$. For the selected targets we obtained high-resolution spectra with the VIMOS High-Resolution red grism (with the GG475 filter, $R\sim2,500$). This grism covers the $3400-4600$ \AA\ rest-frame at the redshift of the target superstructure. The observational configuration of the survey was done so we could measure spectral features such as [O{\sc ii}]\,$\lambda$3726,$\lambda$3729 (partially resolved doublet), the $4000$ \AA\ break, and H$\delta$ for the superstructure members. We have compared our spectroscopic sample with a mass-complete catalogue and correct for sample incompleteness following the procedure detailed in Section \ref{ssection:completeness}.

We measure the redshifts of our sources using \textsc{SpecPro} \citep{masters2011} on the extracted 1D spectra and using a combination of [O{\sc ii}], H+K absorption, G-band, some Fe lines, and H$\delta$. All spectra were visually inspected for the features aforementioned. We obtain secure spectroscopic redshifts for 696 sources with high S/N, of which 490 are at $0.8<z<0.9$. Spectroscopic failures are related to either low S/N continuum or the lack of apparent features.

With the knowledge of the spectroscopic redshift, we improve the estimate of physical parameters that are available in the COSMOS2015 photometric catalogue \citep{laigle2016}. We ran \textsc{magphys} \citep{cunha2008} with spectral models constructed from the stellar libraries by \citet{bruzual2003} and using photometric bands from near-UV to near-IR. The dust is modelled following the prescription described by \citet{charlot2000}. We obtain estimates for the stellar masses and star formation rates for 466 out of the 490 galaxies that are observed at $0.8<z<0.9$. Galaxies with no estimates are mostly fainter \textit{i}-band secondary sources with no match in COSMOS2015 or are not included in the latter catalogue due to different selection bands. Comparing our spectroscopic redshifts with the photometric redshifts of the COSMOS2015 catalogue we find a dispersion $\sigma_{\Delta z/(1+z)}=0.009$. The scatter in stellar mass and SFR  for galaxies with $|z_\mathrm{spec} - z_\mathrm{phot}|<0.1$ is $\sim$0.15 dex and $\sim$0.6 dex respectively. For this study, we use SED derived SFRs since the observed [O{\sc ii}] emission is a poor tracer of SFR for red, low SFR galaxies (with non-detections in quiescent sources) and depends on gas-phase metallicity \citep[e.g.][]{kewley2004, yan2006, kocevski2011} and we have no independent way to quantify dust extinction or measure metallicity from our spectral coverage. We note that there are inevitably issues and limitations affecting SED derived SFRs \citep[especially when young stellar populations dominate galaxies, see e.g.][]{wuyts2011b}, but those affect mostly high SFR galaxies ($>50\mathrm{M_\odot yr^{-1}}$), which is a small portion ($<13$\%) of our sample. We refer to \citeAPA\ for a comparison between SED and [O{\sc ii}] derived SFRs for our sample.

Our final sample is restricted to be at $0.8<z<0.9$ to match our primary selection (see \citeAPA) and has a total of 466 galaxies spanning a large range of environments across several Mpc. We show in Fig. \ref{fig:mass_sfr_dist} the colour-magnitude diagram ($r-z$ versus $z$-band, corresponding roughly to rest-frame $U-V$ vs. $V$) and the stellar mass-SFR relation from our sample and compare it to the parent photometric catalogue. \add{We note that our sample becomes close to a stellar-mass complete sample at stellar masses greater than \msun{10}.}

To estimate the local galaxy over-density we use the density fields computed by \citet{darvish2015a,darvish2017}. These results are based on the photometric redshift catalogue in the COSMOS field provided by \citet{Ilbert2013}. The density field was calculated for a $\sim1.8\,\mathrm{deg^{2}}$ area using a mass-limited ($\log_{10}\left(M/\mathrm{M_\odot}\right)>9.6$) sample at $0.1<z_\mathrm{phot}<1.2$. In this paper we define as over-density

\begin{equation}
1+\delta = \frac{\Sigma}{\Sigma_\mathrm{median}}
\label{eq:density}
\end{equation}

\noindent
with $\Sigma_\mathrm{median}$ being the median of the density field of a specific redshift slice. We used an adaptive kernel with variable size, with small kernel size for crowded regions and larger kernel size for sparser regions, around a typical width of 0.5 Mpc \citep[characteristic size of X-ray clusters, see e.g.][]{finoguenov2007}. We note that re-computing the density field using the spectroscopic redshifts from our sample does not significantly change the underlying density fields. For a more detailed description of the method, we refer to \citet{darvish2015a,darvish2017}. Having a pure density-based definition of the environment does not translate exactly into different physical regions \citep[see e.g.][]{aragon-calvo2010,darvish2014}. In each density bin we can have a mix of different regions, i.e. dense filamentary structures can have similar local densities as the core of rich galaxy groups or cluster outskirts and small groups could share the same local density as scarcely populated filaments. Nonetheless there are typical densities at which field ($\log_{10}(1+\delta)\lesssim 0.1$), filament ($0.1\lesssim \log_{10}(1+\delta)\lesssim 0.6$), and cluster galaxies ($\log_{10}(1+\delta) \gtrsim 0.6$) dominate the population (see \citeAPA\ for more details).

\subsection{Completeness corrections}\label{ssection:completeness}

\begin{figure*}
\centering
\includegraphics[width=\linewidth]{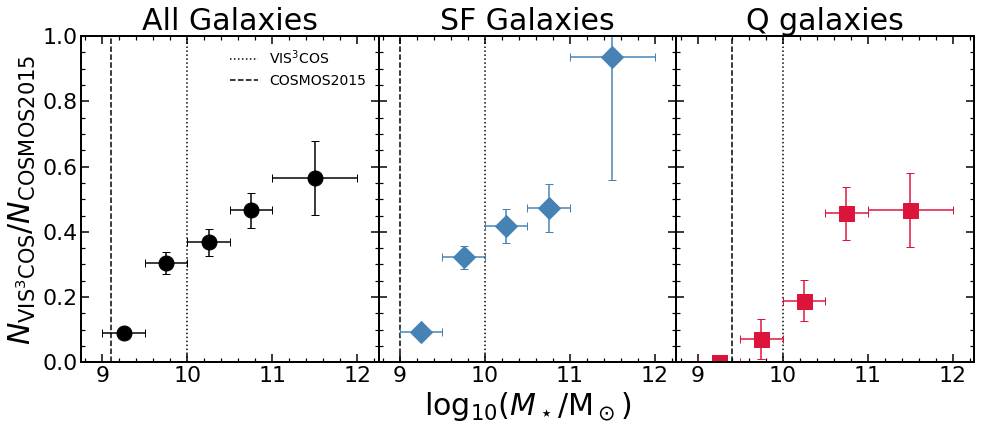}
\caption{\add{The number of galaxies with spectroscopic redshifts in \viscos\ divided by the number of galaxies in the mass-complete COSMOS2015 catalogue over the same region and limited to $0.8<z_\mathrm{phot}<0.9$ for all, star-forming, and quiescent galaxies, respectively. The vertical dashed lines correspond to the mass-complete limits in COSMOS2015. The vertical dotted lines correspond to our quoted limiting mass of $\sim$\msun{10}. The error bars are computed from Poisson statistics.}}
\label{fig:completeness}
\end{figure*}

\begin{figure*}
\centering
\includegraphics[width=0.49\linewidth]{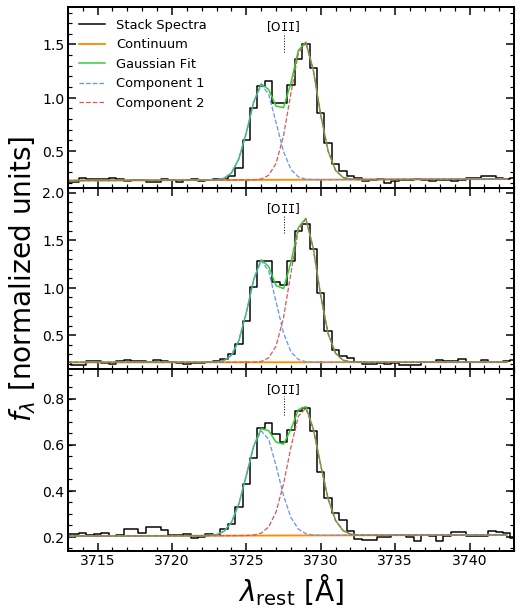}
\includegraphics[width=0.49\linewidth]{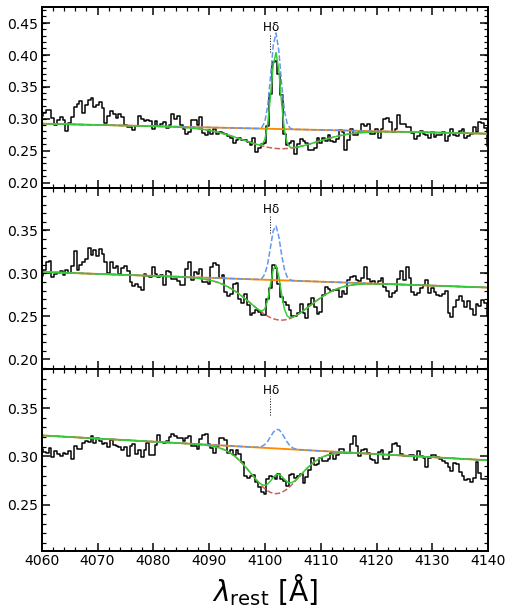}
\caption{Three examples of the fit to the stacked [O{\sc ii}] emission (left) and the stacked H$\delta$ absorption+emission (right) spectral lines. The solid black line shows the observed spectrum. The green line shows the median fit (after 10,000 realizations) and the orange line the estimated continuum around the line. We also show in red and blue dashed lines the fit of each Gaussian component.}
\label{fig:specFitExample}
\end{figure*}

Due to the nature of spectroscopic surveys, it is hard to observe all galaxies in a field down to a single magnitude or stellar mass limit. To make sure that our final spectroscopic sample is as representative as possible of the global population over the same field, we use the excellent wealth of data available in the COSMOS field to quantify the representativeness of our sample. Specifically, we make use of the COSMOS2015 catalogue which is reported to be mass-complete down to \msun{9.4} for quiescent galaxies at these redshifts \citep[mass-completeness limit is \msun{9} if considering all galaxies in the observed region][]{laigle2016}. \add{In Fig. \ref{fig:completeness} we show the relative fraction of galaxies in our spectroscopic sample compared to the mass-complete sample from the COSMOS2015 catalogue. When splitting the samples into the star-forming and quiescent populations, we find different incompleteness effects. Our completeness for quiescent galaxies starts to drop at stellar masses below \msun{10.5}, and we fail to detect a single quiescent galaxy in the lower stellar mass bin (\msun{9}-\msun{9.5}). For star-forming galaxies our completeness drops at lower stellar masses (below \msun{9.5}).}

We assign weights to all of our galaxies in our sample based on the position of the stellar-mass, specific SFR and local over-density with respect to the mass-complete population at $0.8<z_\mathrm{phot}<0.9$. In practice, we compute the fraction of galaxies observed in the three-dimensional region centred on each target in the mass-complete catalogue and compare that to the fraction of galaxies in our spectroscopic sample over the same region:

\begin{equation}
w_i = \frac{N_{\mathrm{mass},R_i}}{N_{\mathrm{spec},R_i}} \times \frac{N_\mathrm{spec}}{N_\mathrm{mass}},
\end{equation}

\noindent
where $N_\mathrm{mass}$ represents the numbers of galaxies in the complete mass catalogue, $N_\mathrm{spec}$ is the number of galaxies in the spectroscopic catalogue, and the index $R_i$ means the variable is computed inside the region surrounding galaxy $i$. To define each region we use a ellipsoidal selection on the three-dimensional space with a size of 0.25 in the $\log_{10}(M_\star/\mathrm{M_\odot})$ dimension, 1 in the  $\log_{10}(sSFR)$  dimension and 0.5 in the $\log_{10}(1+\delta)$ dimension.

We expect these corrections to be valid for all galaxies above \msun{9} which is the quoted mass-completeness limit for star-forming galaxies at these redshifts \citep[][]{laigle2016}. However, since our primary selection would translate to a limiting mass (considering all galaxies) of approximately \msun{10} (this is the median stellar mass of all $22.3<i_\mathrm{AB}<22.7$ and $0.8<z_\mathrm{phot}<0.9$ galaxies in the COSMOS2015 catalogue) \add{and that is reflected in the low completeness values at those stellar masses (see Fig. \ref{fig:completeness}), we restrict our analysis to} galaxies above that stellar-mass threshold. Considering the sub-populations of star-forming galaxies, our primary \textit{i}-band selection translates in different limiting stellar masses. For the star-forming population, our $i-$band selection would typically select galaxies down to \msun{9.9}. For the quiescent population, the same selection would be limited at \msun{10.6}. We try to account for this effect by weighting galaxies based on their position in the stellar mass - sSFR plane, which aims to minimize this selection bias.


\section{Spectroscopic properties}\label{section:methods}

\subsection{Composite spectra}\label{ssection:stacks}

Co-adding spectra of galaxies binned by similar physical properties increase the signal-to-noise ratio and allow for a better determination of the median spectral properties of the sample in different regions of the parameter space that we aim to probe \citep[e.g.][]{lemaux2010}. We construct composite spectra by binning in stellar mass, over-density, and star formation rate. For a quick view on the different composite spectra, see Figs. \ref{fig:stack_mass_full}-\ref{fig:stack_sfr_full}.

To obtain composite spectra, we first normalize each spectrum to the mean flux measured in the range 4150-4350 \AA. Using the redshift we have measured (see Section \ref{section:data}), we linearly interpolate the spectrum onto a common universal grid ($3250-4500$ \AA, $\Delta\lambda=0.5$\,\AA\ $\mathrm{pix^{-1}}$). We then compute the median flux at each wavelength in the grid to obtain the final composite spectra. We normalize each spectrum dividing it by the median flux measured at $4150-4350$ \AA. We assign a weight to each spectrum based on the completeness corrections detailed in Section \ref{ssection:completeness}. Finally, we obtain the composite spectra as the weighted median flux per wavelength bin in the defined grid. We have repeated our analysis using different normalization schemes (blueward of 4000 \AA\ and with no normalization), and our results are qualitatively the same.

\subsection{Spectral quantities}\label{ssection:linemeasures}

To study the star formation history of galaxies, we use three tracers - [O{\sc ii}], H$\delta$, $D_n4000$ -  present in our spectra \citep[e.g.][]{balogh1999, dressler2004, oemler2009, poggianti2009, vergani2010, mansheim2017}. To be consistent with the classical notation, a negative equivalent width corresponds to a line in emission while positive values correspond to a line in absorption.

\subsubsection{[O{\sc ii}] emission}

To measure the [O{\sc ii}], we fit a double Gaussian to the doublet. The centre of each component is set to be $\lambda_1=3726.08\pm0.3$\AA\ and $\lambda_2=3728.88\pm0.3$\AA\ (a small shift in the line centre is allowed to account for our finite resolution, and we allow for a systematic shift to the doublet to account for redshift uncertainties). We measure the flux and line equivalent widths by integrating over the best-fit models. For more details we refer the reader to \citeAPA\ (see also Fig. \ref{fig:specFitExample}, left panel).

\subsubsection{H$\delta$ in emission and absorption}

We fit the emission and absorption components of the H$\delta$ line by using a double Gaussian fit with two independent components (see Fig. \ref{fig:specFitExample}, right panel): one forced to have a negative amplitude (for the absorption) and one forced to have a positive amplitude (for the emission). To prevent a set of degenerate model combinations which produce the same combined result but for which the individual components are clearly not physically representative of the observed data \footnotemark\footnotetext{The sum of two symmetric components is degenerate against an equal multiplicative factor on the amplitude of individual components.} we force the width of the emission line to be always smaller than the absorption component. This is informed by the empirical information from the individual stacks we obtain. To estimate the first guess amplitude of the absorption, we take the minimum of the continuum subtracted spectra. The amplitude of the emission line is computed by first fitting a single Gaussian to the absorption feature (whenever present) masking the $\pm3$ \AA\ around the central wavelength, and then taking the maximum of the absorption subtracted spectra. We also assign an initial $\sigma$ of 4 and 1 \AA\ for the absorption and emission component, respectively. We then measure the line fluxes and equivalent widths from the best-fit models.

\subsubsection{Continuum and error estimation}\label{sssec:errors}

For all line fits we individually define two regions (one blue-ward, one red-ward of the line) with a width of 15\,\AA\ width ($\sim$3 times the spectral resolution) from which we estimate the median continuum level. Then the local continuum is defined as a straight line that goes through those two points \citep[see e.g.][]{balogh1999,lemaux2010,mansheim2017}. To minimize the effect of a particular choice of windows from which we compute the continuum, we compute the median flux for 5,000 random shifts of $k$\AA\ on the proposed interval, where $k$ is randomly drawn from a normal distribution centred at 0 and with a width of 5 \AA. Then the final estimate of the continuum is measured from the median of the 5,000 realizations.

To estimate the errors on the derived spectral quantities, we have performed a bootstrap sampling of each sub-sample of spectra. This method allows us to estimate the variance occurring within each sub-sample. This quoted error is always larger than the formal uncertainties from the fit. We perform the fit on 1,000 individual bootstrapped spectra (using only 80\% random galaxies drawn from the sub-sample) and derived the final errors from the 16th and 84th percentiles of the distribution of best-fit values. 

\subsubsection{$4000$ \AA\ break}

Apart from line measurements, we have also computed the strength of the break at 4000v\AA\ \citep[$D4000$ and $D_n4000$ defined by ][, respectively]{bruzual1983,balogh1999}. We automated the computation of these quantities by integrating the spectra over the red ($D4000$: 4050-4250 \AA, $D_n4000$: 4000-4100 \AA) and blue ($D4000$: 3750-3950 \AA, $D_n4000$: 3850-3950 \AA) intervals and computing the ratio of those fluxes as

\begin{equation}
X4000 = \frac{\int_{\lambda_{r1}}^{\lambda_{r2}} f_\nu d\lambda}{\int_{\lambda_{b1}}^{\lambda_{b2}} f_\nu d\lambda},
\end{equation}

\noindent
where $X$ is either $D$ or $D_n$ depending on the integration limits of the red ($\lambda_{ri}$) and blue ($\lambda_{bi}$) intervals. When comparing both indices, we find them to correlate well, with a median difference of $<1\%$ and a spread of $30\%$ on individual measurements. We opt to use for the remainder of the paper the value of $D_n4000$ since it should be less affected by errors due to Poisson sampling and less affected by reddening \citep{balogh1999}. Nevertheless, our results are qualitatively the same regardless of which index we use. To avoid the contamination by emission lines in the integrated regions we mask 6 \AA\ regions around the [Ne{\sc iii}] and H$\zeta$ lines (see e.g. Fig. \ref{fig:stack_mass_full}).

\subsection{Stellar population age estimates}\label{ssection:stellar_age}

\begin{figure}
\centering
\includegraphics[width=\linewidth]{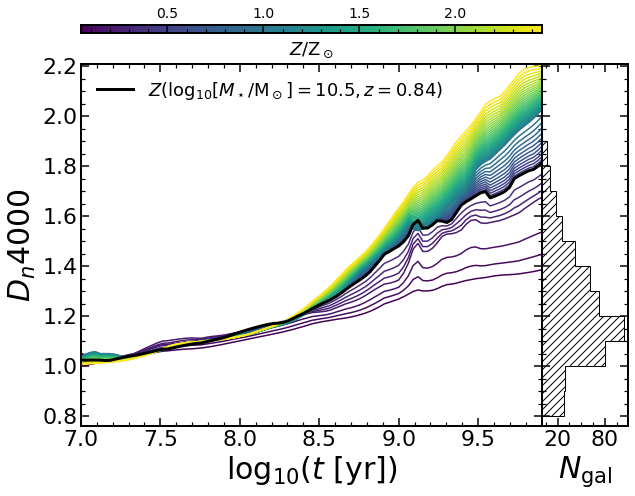}
\caption{The value of $D_n4000$ as a function of stellar age for a set of single stellar population models from \citet{bruzual2003} for different stellar metallicities from $Z=0.1\mathrm{Z_\odot}$ to $Z=2.5\mathrm{Z_\odot}$. The thick solid line shows the relation using the stellar metallicity for the median stellar mass and redshift of our sample (see Section \ref{ssection:stellar_age}). On the right panel, we show the distribution of measured $D_n4000$ values for the galaxies in our sample.}
\label{fig:d4000_Age} 
\end{figure}

Estimating a single representative stellar age for galaxies is not a trivial task due to the expected variety of their star formation histories. Nonetheless, we can obtain an estimate given a few sets of assumptions. We will use the $D_n4000$ index as a proxy for age and obtain an estimate from a set of stellar population models described by \citet{bruzual2003}. We attempt to estimate an age based on a single stellar population (SSP), with the notation $t_\mathrm{SSP}$, which should trace the age of the last major burst that the galaxy had. 

We note that the $D_n4000$ index depends not only on age but also on the stellar metallicity, especially for ages greater than 1 Gyr \citep[e.g.][]{bruzual1983,poggianti1997,balogh1999}. Since we do not have any independent way to estimate stellar metallicity for all galaxies, we need to make a few assumptions to try and mitigate possible bias in our interpretations. We assume that our sample follows the stellar mass-metallicity relation that is found locally and up to $z\sim1$ \citep[e.g.][]{tremonti2004,gallazzi2005,savaglio2005,zahid2011,zahid2013,ma2016,derossi2017,leethochawalit2018}. We then estimate individual stellar metallicities based on a recent numerical simulation study by \citet{ma2016} which parametrizes stellar metallicity as a function of stellar mass and redshift as:

\begin{equation}
\log_{10}\left(\frac{Z_\star}{\mathrm{Z_\odot}}\right) = 0.40\left[\log_{10}\left(\frac{M_\star}{\mathrm{M_\odot}}\right) - 10\right] +0.67e^{-0.5z}-1.04.
\end{equation}

\noindent
For a stellar mass ($10^{10.5}\mathrm{M_\odot}$) and redshift ($z=0.84$) of our sample we determine a stellar metallicity of $Z_\star \sim 0.008 = Z_\odot/2.5$. For a stellar mass of $10^{11.5}\mathrm{M_\odot}$ we get $Z_\star \sim 0.02$ and for a stellar mass  $10^{9.5}\mathrm{M_\odot}$ we get $Z_\star \sim 0.003$. We show in Fig. \ref{fig:d4000_Age} the age dependence of $D_n4000$ for different stellar metallicities spanning the expected range of our sample.

To estimate the age of a galaxy (or group of galaxies) we first compute the individual (or median) stellar metallicity and then use the $D_n4000$-age relation for an SSP based on \citet{bruzual2003} models at that metallicity to derive the stellar age. The errors on the SSP ages are derived as detailed in Section \ref{sssec:errors} and do not include any uncertainty on the metallicity.


\section{Results}\label{section:results}

We explore the results on the spectral properties of our galaxies at $0.8<z<0.9$, highlighting both composite spectra and individual galaxies in this section. We aim to probe the influence of key physical properties (stellar mass, environment, and SFR) on the median observed spectral properties of our sample. We note that for H$\delta$ we have insufficient S/N on most galaxies to get robust measurements and we refrain from discussing that spectral feature in terms of individual galaxies.

\begin{figure*}
\centering
\includegraphics[width=0.85\linewidth]{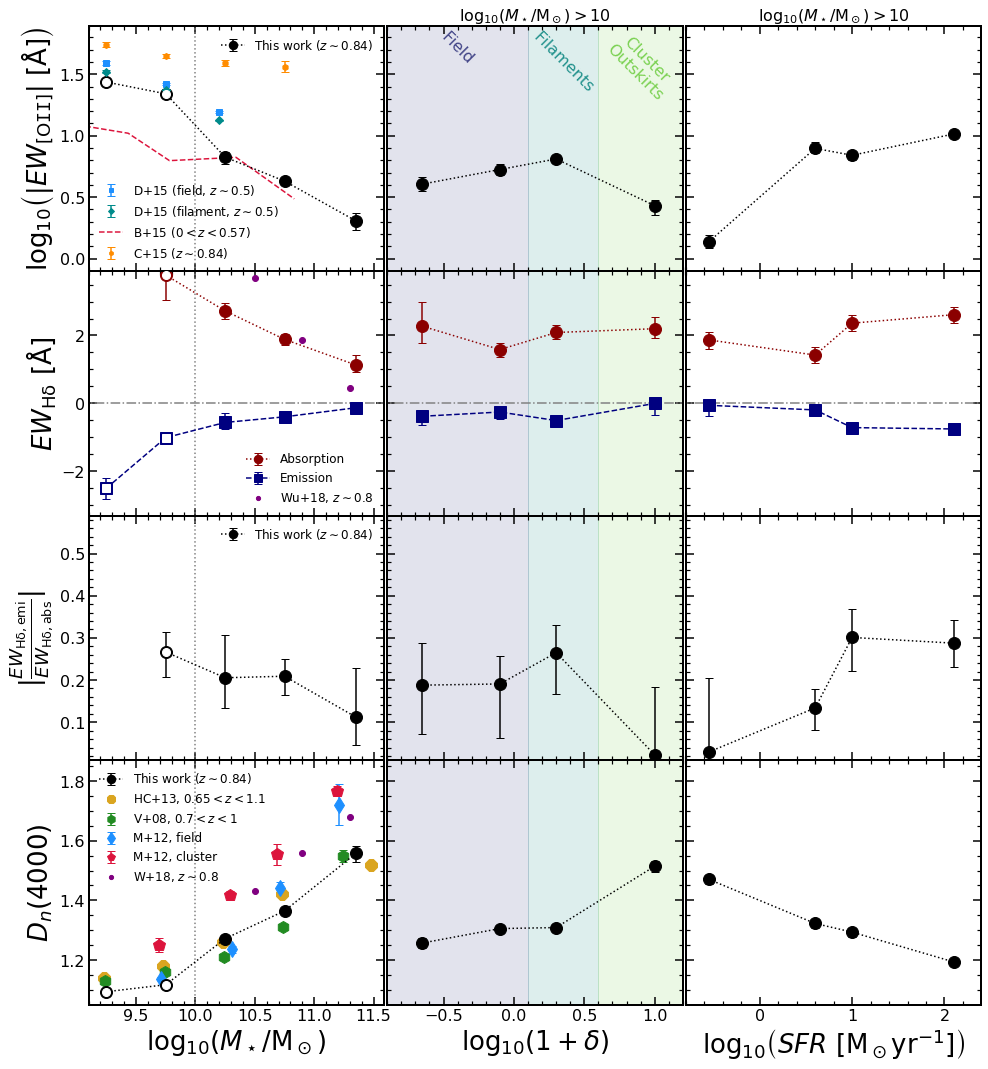}
\caption{The dependence of the three spectral features detailed in Section \ref{ssection:linemeasures} (from top to bottom - [O{\sc ii}] equivalent width, H$\delta$ equivalent widths, and $D_n4000$) as a function of stellar mass (left), over-density (middle), and SED-derived SFR (right). We show our stellar mass representativeness limit as a vertical dotted line in the left panels \add{and show as empty symbols the measurements for which the completeness effects are large.} We note that for bins not defined in stellar mass (middle and right panels), the composite spectra are built from galaxies with stellar mass greater than $10^{10}\mathrm{M_\odot}$. Results from this paper are shown as large and dotted line connected symbols with associated error bars derived from the 16th and 84th percentiles of the 1,000 bootstrapped fits (if the error bars are not seen, it implies an error smaller than the symbol size). When looking at trends with stellar mass, we see a decrease in [O{\sc ii}] and H$\delta$ emission (blue squares) and absorption (red circles) strength and an increase of the average age of the stellar population (traced by $D_n4000$). For trends with over-density, there is a peak in [O{\sc ii}] equivalent width at filament-like densities and H$\delta$ equivalent widths show little dependence on local density. There is also a clear trend for galaxies being older in higher density regions. Lastly, both [O{\sc ii}] and H$\delta$ strength increase with SFR (excluding the most star-forming galaxies) and we see younger populations in galaxies with higher SFR, as expected. We compare our results to other surveys in the literature \citep{bridge2015, darvish2015, cava2015, wu2018, hernan-caballero2013, vergani2008, muzzin2012}. For a more detailed discussion on the differences we refer to Sections \ref{ssection:oii} and \ref{ssection:d4000}.}
\label{fig:stack_properties}
\end{figure*}

We summarize in Fig. \ref{fig:stack_properties} (see also Appendix \ref{app:stack_results} and Table \ref{tab:results}) the properties of composite spectra on [O{\sc ii}] and H$\delta$ (both emission and absorption) line equivalent widths and $D_n4000$ at different stellar masses, over-densities, and SFRs. We stress that for samples not selected in stellar mass, we impose a minimum stellar mass limit of $10^{10}\mathrm{M_\odot}$.

\subsection{Global trends on the spectroscopic properties}

\subsubsection{Stellar mass}\label{ssection:results_mass}

\add{We discuss here the trends of the measured spectroscopic properties as a function of the stellar mass. While we show in Fig. \ref{fig:stack_properties} the results at stellar masses between \msun{9} and \msun{10}, we do not discuss them as this results suffer from high incompleteness (see Section \ref{ssection:completeness}).} 

In terms of the [O{\sc ii}] line equivalent width (EW$_\mathrm{[O\mathtt{II}]}$), we find in Fig. \ref{fig:stack_properties} a strong decrease with stellar mass, \add{by a factor of $\sim3$ in strength from the lowest ($10<\log_{10}\left(M_\star/\mathrm{M_\odot}\right)<10.5$)} to the highest stellar mass bin ($\log_{10}\left(M_\star/\mathrm{M_\odot}\right)>11$), which points to a decrease in sSFR with increasing stellar mass from $\sim 10^{-9} \mathrm{yr^{-1}}$ to $\sim 10^{-10} \mathrm{yr^{-1}}$ \citep[a consequence of the main sequence of star-forming galaxies, see also][]{darvish2015}. We compare the results for individual galaxies and composite spectra with others available in the literature \citep{bridge2015, cava2015, darvish2015}. Our results are broadly consistent (in terms of the observed trends) with the literature which find a decrease in the absolute line equivalent width with increasing stellar mass. However, we find some discrepancies with \citet{cava2015} and \citet{bridge2015} in terms of the average value in bins of stellar mass that are likely related to the target selection in each work. In the case of \cite{cava2015} they report consistently higher values of [O{\sc ii}] line equivalent width. However, their sources are selected through medium band filters down to equivalent widths of $\sim$15-20 \AA, which naturally explains their higher median values. As for the case of \citet{bridge2015} they study a large field with blind spectroscopy, and they exclude large equivalent width ($EW \gtrsim 40$ \AA) galaxies to avoid contamination by higher redshift interlopers (Ly$\alpha$ emitters), which can explain their observed lower equivalent widths. We find $EW_\mathrm{[O\mathtt{II}]}$ values consistent with \citet{darvish2015}, which had a similar observational setup as the VIS$^{3}$COS survey.

In the middle panels of Fig. \ref{fig:stack_properties} we show the measured equivalent width of the H$\delta$ emission and absorption for all composite spectra which allowed a measurement (some of them did not show any signs of absorption). The H$\delta$ absorption line equivalent width decreases with increasing stellar mass \add{(from EW$_\mathrm{H\delta} = 2.7\pm0.2$ \AA~at $10<\log_{10}\left(M_\star/\mathrm{M_\odot}\right)<10.5$ down to EW$_\mathrm{H\delta} = 1.1_{-0.2}^{+0.3}$ \AA\ at $\log_{10}\left(M_\star/\mathrm{M_\odot}\right)>11$)}. The H$\delta$ emission line equivalent width also correlates with stellar mass with stronger emission being found at lower stellar masses, \add{dropping from EW$_\mathrm{H\delta} = -0.6_{-0.2}^{0.3}$ \AA\ at $10<\log_{10}\left(M_\star/\mathrm{M_\odot}\right)<10.5$ to a marginally non-existent emission component with EW$_\mathrm{H\delta} = -0.1 \pm 0.1$ \AA\ at $\log_{10}\left(M_\star/\mathrm{M_\odot}\right)>11$}. \add{When compared to the results by \citet{wu2018} we see that there is also a decrease on their reported equivalent width with stellar mass but at a steeper rate. We attribute the discrepancies to the different methods used to compute the line equivalent width. They use a spectral index defined by \citet{worthey1997} which measures the line equivalent width on an emission subtracted/masked spectrum.} We also show the dependence of the emission to absorption ratio of the equivalent widths as a proxy for the ratio of O to A stars. Our results are consistent with no dependence on stellar mass.

The bottom panels of Fig. \ref{fig:stack_properties} show the dependence of the 4000 \AA\ break strength on the same quantities mentioned above. We find a strong correlation between $D_n4000$ and stellar mass, \add{increasing from $D_n4000=1.27_{-0.01}^{+0.02}$ at $10<\log_{10}\left(M_\star/\mathrm{M_\odot}\right)<10.5$} to $D_n4000=1.56\pm0.03$ at $\log_{10}\left(M_\star/\mathrm{M_\odot}\right)>11$. This trend points to more massive galaxies also having the older stellar populations. Our results are in general agreement with other studies in the literature targeting either clustered regions \citep[GCLASS -][]{muzzin2012} or large field surveys \citep[VVDS, SHARDS, LEGA-C - ][\!\!, respectively]{vergani2008,hernan-caballero2013,wu2018}. The trend with stellar mass is seen in all surveys. We find median values in between the quoted average/median of other studies in the literature at similar redshifts. We find lower median $D_n4000$ at fixed stellar masses with respect to values reported by \citet{muzzin2012} for cluster galaxies, which is expected given the dependence of $D_n4000$ seen with the environment and the fact that we are probing a majority population at lower densities than the cluster sample. This is consistent with what we find when splitting the sample in stellar mass and local density bins, with galaxies at fixed stellar mass having the stronger 4000 \AA\ breaks at the higher densities we probe (see Section \ref{ssection:massAndEnvironment}).

\subsubsection{Local environment}\label{ssection:results_env}

Our analysis is restricted to stellar masses greater than $10^{10}\mathrm{M_\odot}$ and for this sub-sample there is little variation in the median stellar mass across the different bins ($\Delta\log_{10}\left(M_\star/\mathrm{M_\odot}\right)<0.15$). We find that the absolute value of EW$_\mathrm{[OII]}$ increases from $4.0^{+0.5}_{-0.6}$ \AA\ to $6.5^{+0.4}_{-0.5}$ \AA\ from field to filament-like densities and then drops to $2.7^{+0.5}_{-0.3}$ \AA\ in our highest density bin. This is pointing to a slight increase of the sSFR at filament-like densities for galaxies more massive than $10^{10}\mathrm{M_\odot}$ and then a strong drop towards the higher density regions probed here. We also compared the trends with local density to those with environmental regions (field, filament, and cluster, see Table \ref{tab:results}) as defined in \citeAPA\ \citep[see also][]{darvish2015a,darvish2017} and find a decline of |\EWoii\!| from field to filament to cluster regions (from $4.6^{+0.5}_{-0.4}$ \AA\ to $4.4^{+0.3}_{-0.4}$ \AA\ to $3.4\pm0.5$ \AA, respectively), for galaxies more massive than $10^{10}\mathrm{M_\odot}$. These differences reflect the nuances of using different tracers of the galactic environment, but both reinforce a drop of  \EWoii towards the denser regions studied here.
 
Regarding the H$\delta$ absorption line, we find results that are consistent with no trend with local density. For H$\delta$ emission, all our derived values for galaxies more massive than $10^{10}\mathrm{M_\odot}$ are also consistent with no dependence with over-density, having measured equivalent widths around $\sim$0-0.5 \AA. We find no significant dependence on over-density on the results regarding the ratio between absorption and emission of H$\delta$, given our error bars.

Finally, we find an increase of $D_n4000$ towards higher densities. In low- to intermediate-density regions ($\log_{10}(1+\delta)<0.5$) we find $D_n4000 \sim 1.26-1.31$ (corresponding to an SSP age of $\sim0.35-0.47$ Gyr). The strength of the 4000 \AA\ break then increases at higher densities ($\log_{10}(1+\delta)>0.5$), reaching $D_n4000=1.51\pm0.02$ (corresponding to an SSP age of $\sim1.1\pm0.1$ Gyr).

\subsubsection{Star formation rate}\label{ssection:results_sfr}

Our analysis is restricted to stellar masses greater than $10^{10}\mathrm{M_\odot}$, which probes a range of 3 dex in SFR. 
\add{We note that the difference in the median/mean stellar mass in each defined SFR bins is smaller than 0.1 dex, so the correlations of the studied tracers with SFR are not affected by an underlying SFR-stellar mass relation. This happens since at stellar masses $>$\msun{10} there is a large spread in star-formation rates due to existing star-forming and quiescent galaxies of similar stellar mass. 
} 
\add{Overall, there is an increase of \EWoii with increasing SFR. We find a strong increase from the quiescent (low-SFR) population ($-1.4\pm0.2$ \AA\ for $\log_{10}(SFR)<0.4$) to the active (intermediate-SFR) star-forming population ($\sim$-7.5 \AA) in our sample. Then we find a small increase towards the high-SFR population ($-10.3\pm0.5$ \AA\ for$\log_{10}(SFR)>1.2$).}
 
Concerning the H$\delta$ absorption component, we find a small increase of the equivalent width with SFR, from EW$_\mathrm{H\delta} \approx 1.9-1.4 \pm 0.25$ \AA\ at $\log_{10}(\mathrm{SFR}\ \mathrm{[M_\odot yr^{-1}]})<0.8$ to EW$_\mathrm{H\delta}  \approx 2.4-2.6 \pm 0.25$ \AA\ at $\log_{10}(\mathrm{SFR}\ \mathrm{[M_\odot yr^{-1}]})>0.8$. The H$\delta$ emission shows a correlation with SFR that mirrors what we find with [O{\sc ii}] emission with the line strength increasing from low to high SFRs (from $-0.1_{-0.3}^{+0.1}$ \AA\ to $-0.8\pm0.1$ \AA). The emission to absorption line ratio is marginally consistent with no trend with SFR (though we see a rise on the median value from low to intermediate-high SFR galaxies).

Finally, we observe a steady decrease in the value of $D_n4000$ from the lowest SFR bin ($D_n4000=1.47\pm0.01$ for $\log_{10}(SFR\ \mathrm{[M_\odot yr^{-1}]})<0.4$) to the highest SFR bin ($D_n4000=1.19\pm0.01$ for $\log_{10}(SFR\ \mathrm{[M_\odot yr^{-1}]})>1.2$). This is consistent with what is expected from the evolution of galaxies, since we expect a larger fraction of young stars in highly star-forming galaxies which decreases the value of $D_n4000$.

\subsection{Disentangling environment and stellar mass effects}\label{ssection:massAndEnvironment}

To disentangle the effects of stellar mass and environment, we have explored our sample binned both in stellar mass and over-density bins. We chose the over-density bins in a way that they should be representative of field (low-density), filament (intermediate density), and cluster outskirts (high-density) regions (see \citeAPA\ for more details). We show for individual galaxies and composite spectra the influence of over-density in the observed spectroscopic properties in three different bins of stellar mass (chosen as a compromise for reasonable S/N for H$\delta$). 

\subsubsection{EW$_\mathrm{[OII]}$}\label{ssection:oii}

\begin{figure}
\centering
\includegraphics[width=\linewidth]{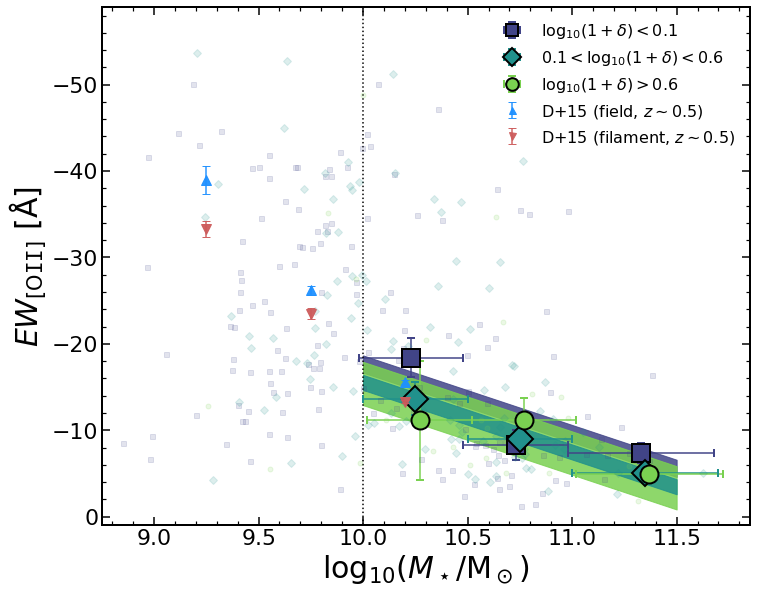}
\caption{We show the relation between [O{\sc ii}] equivalent width and stellar mass for the [O{\sc ii}]-emitters in our sample in three different bins of over-density. We compare our results (large green and dark blue symbols are the median of the population, and the same smaller symbols represent individual measurements) with results from \citet{darvish2015} of filament/field galaxies, as small red/blue triangles, respectively. We show the best fit (with error estimate) from Eq. \ref{eq:oii_mass} for each density bin as shaded regions. We show that higher stellar mass galaxies have weaker [O{\sc ii}] emission and this relation is seen in all over-density subsets.}
\label{fig:OII_Mass_literature} 
\end{figure}

We show in Fig. \ref{fig:OII_Mass_literature} the relation between [O{\sc ii}] line equivalent width and stellar mass for individual galaxies {with an [O{\sc ii}] detection, which should \add{mainly trace star-forming galaxies, though we find a $\sim$10\% contamination (17 out of 166  [O{\sc ii}] emitters) in our $>$\msun{10} sample from low sSFR ($\log_{10}(sSFR)<-11$) galaxies} \citep[see also e.g.][]{yan2006,lemaux2010}. We attempt to separate the effects of local density on this correlation and find that the correlation between [O{\sc ii}] line equivalent width and stellar mass is similar in all environments. We find similar gradients at all environments and fit a linear relation with a fixed slope (the average of individually fitted slopes)\footnotemark{}\footnotetext{We made this choice since individual values for the slope are found to be similar and within the reported errors, and by doing so we can report the change in normalization independent of the slope of the relation.} for all environments:

\begin{equation}
EW_\mathrm{[OII]}  = 8.05\times\log_{10}\left(M_\star/10^{10}\mathrm{M_\odot}\right) + b.
\label{eq:oii_mass}
\end{equation}  

\noindent
The best fit values are shown in Table \ref{tab:linefits}. Line equivalent widths should be insensitive to dust if both continuum and line flux are being emitted from the same regions. We note, however, that there is a different attenuation of stellar and nebular emission seen in the local Universe \citep[e.g.][]{calzetti2000,wild2011}, which seems to be less pronounced at higher redshifts \citep[$z\gtrsim1$, e.g.][]{kashino2013,pannella2015}. We tentatively find a lower [O{\sc ii}] equivalent width with increasing density, but all relations are within 1$\sigma$ uncertainties. 
This is consistent with no dependence of the sSFR with environment for a star-forming population \cite[e.g.][\!; \citeAPA]{muzzin2012,koyama2013,darvish2016}.

Concerning the results from composite spectra, we show in Fig. \ref{fig:stacksProps_dualMass} an overall trend [O{\sc ii}] line equivalent width with stellar mass (as reported in Fig. \ref{fig:OII_Mass_literature}) at each over-density bin. We show that [O{\sc ii}] depends on both the stellar mass and environment of galaxies. For lower stellar mass galaxies ($10<\log_{10}\left(M_\star/\mathrm{M_\odot}\right)<10.5$) we find a small decrease in EW$_\mathrm{[OII]}$ with increasing local density. Intermediate-mass galaxies ($10.5<\log_{10}\left(M_\star/\mathrm{M_\odot}\right)<11$) show a small rise (at the 1.7$\sigma$ level) in EW$_\mathrm{[OII]}$ from field to filament-like regions and then a small decrease towards cluster-like regions. The most massive galaxies are consistent with no environmental effect on the [O{\sc ii}] emission. The local density dependence of the \EWoii emission in the stacked spectra can be reconciled with the apparent environmentally-independent estimates we show in Fig. \ref{fig:OII_Mass_literature}. We include in the stacking analysis all the galaxies with no [O{\sc ii}] emission \add{(either quiescent or dusty) which are known to be more common in higher density regions \citep[quiescent - e.g.][\!; \citeAPA; dusty star formation - e.g.  \citealt{smail1999,galazzi2009,koyama2013,sobral2016}]{peng2010b, cucciati2010b, sobral2011, muzzin2012, darvish2016}.}

	\subsubsection{$4000$ \AA\ break}\label{ssection:d4000}

\begin{figure*}
\centering
\includegraphics[width=0.49\linewidth]{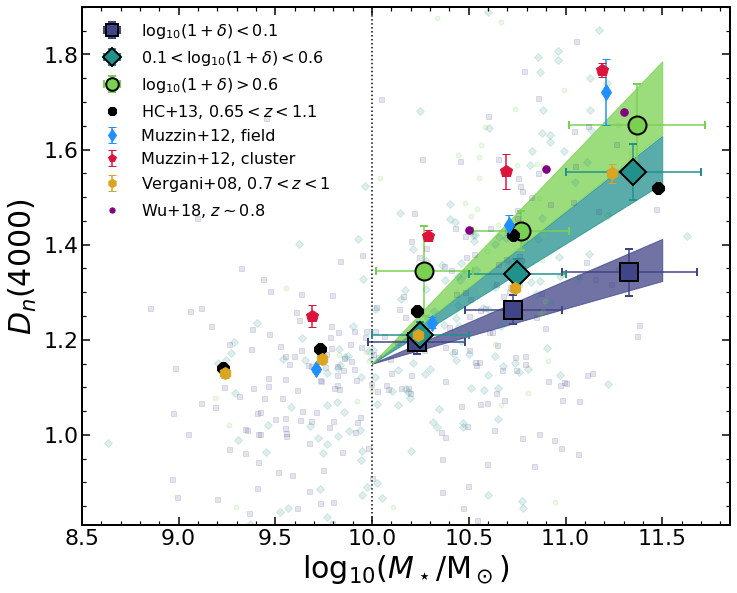}
\includegraphics[width=0.49\linewidth]{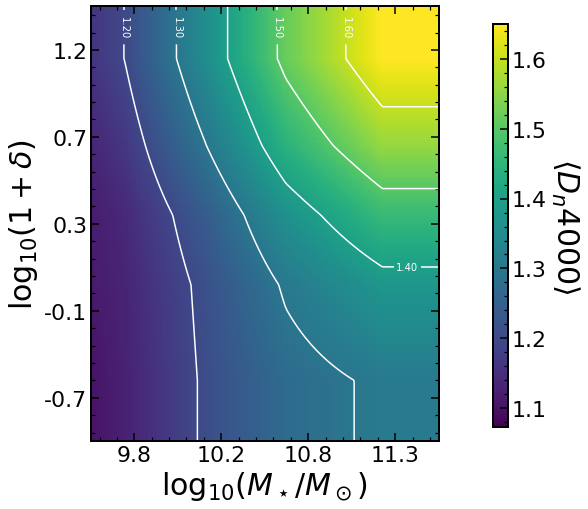}
\caption{$D_n4000$ as a function of stellar mass for three different over-density bins. The vertical dotted line shows the representativeness limit of our survey. We show the best fit (with errors) from Eq. \ref{eq:d4000_mass} for each density bin as shaded regions. We find an underlying correlation between stellar masses and $D_n4000$, with the median $D_n4000$ of galaxies in high-density regions being larger than what is found in low-density regions. The difference between low and high-density environments is larger at higher stellar masses. We compare with the results from GCLASS \citep{muzzin2012}, from VVDS \citep{vergani2008}, from SHARDS \citep{hernan-caballero2013}, and from LEGA-C \citep{wu2018} which also show the same trends. \add{We also show here the median value of  $D_n4000$ as a function of both stellar mass and local density on in the right panel.} This map is a smooth interpolation of the trends found in our sample \add{and highlights the increase of the median $D_n4000$ with increasing stellar mass and local density}.}
\label{fig:D4000_Mass}
\end{figure*}

The 4000 \AA\ break is a proxy for the age of the underlying stellar population (see Section \ref{ssection:stellar_age}). Under that assumption we find that at higher masses ($\log_{10}\left(M_\star/\mathrm{M_\odot}\right)>10$) galaxies residing in high-density regions are typically older than their counterparts at lower density regions. We find that $D_n4000$ is $5\pm3$\%  (lower stellar mass, $\sim$0.3 Gyr difference) to $23\pm6$\% (higher stellar mass, $\sim$2 Gyr difference) higher in high-density regions when compared to lower density regions, see Fig. \ref{fig:D4000_Mass}. There is also an underlying correlation between stellar mass and $D_n4000$, with more massive galaxies having stronger flux breaks at $4000$ \AA, thus being older \citep[rising from $\sim1.1$ to $\sim1.35-1.65$ from lower to higher stellar masses, see also e.g.][]{vergani2008,muzzin2012,hernan-caballero2013,wu2018}. 

The relative difference between field and cluster galaxies shown by \citet{muzzin2012} indicates a stronger break in higher density regions, although, in their sample, the difference between cluster and field galaxies becomes smaller with increasing stellar mass. This can be interpreted as a stronger dependence of the quenched fraction on environment for lower stellar masses \citep[e.g.][]{peng2010b} which is less evident at $z\sim1$ \citep[][though we do find a dependence for our sample, see \citeAPA]{muzzin2012,darvish2016}. However, one also needs to account for the different timescales of the tracer used to defined quiescence (\add{we use the SED-based sSFR, which traces the past $\sim$100 Myr}) and $D_n4000$ which traces the average age of the stellar population (and can range from several hundreds of Myr to several Gyr, see e.g. Section \ref{ssection:quiescentAndSF} where we show that quiescent galaxies in high-density environments have larger $D_n4000$ when compared to the quiescent population in low-density regions).

We also note that the separation between field and cluster galaxies in \citet{muzzin2012} is done by using the cluster-centric radius (defined as the distance to the brightest cluster galaxy of each of their clusters). Their field sample is representative of a population of galaxies in-falling into the clusters, and their cluster galaxies are drawn from a sample of rich clusters. This means that making a direct comparison is not straightforward. Thus we find that their field galaxies should correspond to filament-like densities as defined in our paper and cluster galaxies likely correspond to higher densities than what we probe with the VIS$^{3}$COS survey. Nonetheless, the quiescent fraction that is reported in \citet{muzzin2012} is in line with that found in studies where field samples are independent of cluster regions and do not reach rich cluster density regions \citep[][]{vanderburg2013,peng2010b}. It is therefore not clear that the definition of field and cluster samples can explain the observed differences in the differential trend of $D_n4000$ and stellar mass for field and cluster galaxies.

Quantifying the rate of increase of $D_n4000$ with the linear model (we fix the y-intercept to the average of individual fits)\footnotemark{}\footnotetext{We made this choice since individual values for the y-intercept are within the reported errors, and by doing so we can report the change in gradient independent of the normalization of the relation.}

\begin{equation} 
D_n4000 = m\times\log_{10}\left(M_\star/10^{10}\mathrm{M_\odot}\right) + 1.15,
\label{eq:d4000_mass}
\end{equation}

\noindent
where the best fit values are summarized in Table \ref{tab:linefits}.

\begin{table*}
\centering
\caption{Results of the linear fits $X = m\times\log_{10} (M_\star/10^{10}\mathrm{M_\odot} ) +b$ to EW$_\mathrm{[OII]}$ and $D_n4000$ that are shown in Figs. \ref{fig:OII_Mass_literature} and \ref{fig:D4000_Mass}. We note that the slope is fixed for the [O{\sc ii}] related fits, and the y-intercept is fixed for the $D_n4000$ related fits.}
\label{tab:linefits}
\begin{tabular}{ccccccc}
\hline
{$X$} & \multicolumn{2}{c}{$\log_{10}(1+\delta)<0.1$}	&	\multicolumn{2}{c}{$0.1<\log_{10}(1+\delta)<0.6$} & 	\multicolumn{2}{c}{$\log_{10}(1+\delta)>0.6$} \\
				& $m$ & $b$  & $m$ & $b$ & $m$ & $b$\\
\hline
EW$_\mathrm{[OII]}$ & $[8.04]$ & $-17\pm1$ & $[8.04]$ & $-16\pm1$ & $[8.04]$ & $-15\pm2$ \\
$D_n4000$ & $0.14\pm0.02$ & $[1.15]$ & $0.29\pm0.03$ & $[1.15]$ & $0.37\pm0.05$ & $[1.15]$ \\
\hline
\end{tabular}
\end{table*}

We confirm more clearly the trend reported on individual galaxies when looking at $D_n4000$ in the composite spectra (see Fig. \ref{fig:stacksProps_dualMass}). At fixed density, we see that $D_n4000$ increases from low to high stellar masses. At fixed stellar mass, we find that $D_n4000$ increases from low to high-density environments (with the exception at the highest densities where galaxies between \msun{10} and \msun{11} have similar values of $D_n4000$). We also find that the difference between different stellar mass bins is different for different density regions. The relation becomes steeper at intermediate densities when compared to the low-density regions probed here. At the higher densities we probe, the difference between the low and intermediate stellar mass bins is smaller, but we find a larger difference towards the highest stellar masses. This points to both stellar mass and environment having an impact on the stellar populations of galaxies, with higher density environments harbouring older galaxies at all stellar masses.

\begin{figure}
\centering
\includegraphics[width=\linewidth]{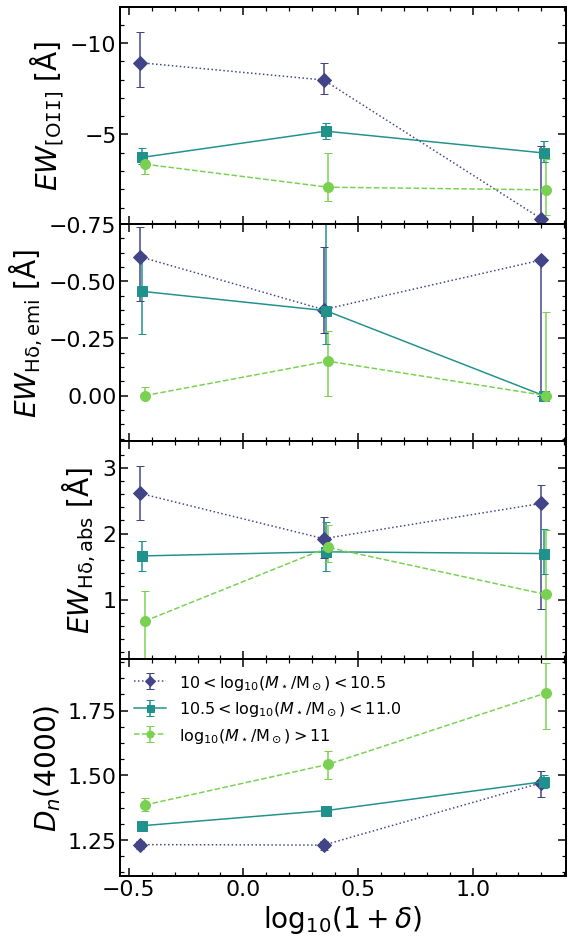}
\caption{The dependence of the three spectral indices detailed in Section \ref{ssection:linemeasures} (from top to bottom - [O{\sc ii}] equivalent width, H$\delta$ equivalent width, and $D_n4000$) as a function of over-density in three stellar mass bins. This highlights the impact of both stellar mass and environment on the observed spectral properties of galaxies.}
\label{fig:stacksProps_dualMass}
\end{figure}

\subsubsection{H$\delta$ emission and absorption}

We also find a dependence on the stellar mass of the strength of the H$\delta$ emission with lower stellar mass galaxies having on average higher equivalent widths (see Fig. \ref{fig:stacksProps_dualMass}, Table \ref{tab:results}). Within our estimated errors, we cannot pinpoint any dependence of the line strength with the environment. Concerning the absorption component of H$\delta$, the results hint at a dependence on the environment that depends itself on the stellar mass we consider. \add{However, only the rise from low to intermediate densities in absorption EW for higher stellar masses is of some significance  (from  $0.7^{+0.5}_{-0.7}$\AA\ to $1.8^{+0.3}_{-0.2}$\AA).} Interestingly, at intermediate densities, the H$\delta$ absorption line has a similar equivalent width at all stellar masses.

\add{With respect to the emission component of H$\delta$, we find a tentative trend of the emission being stronger for less massive galaxies (as reported for the full sample in Fig. \ref{fig:stack_properties}). We find no significant dependence with local density for each of the stellar mass ranges considered.}

\subsubsection{Anti-correlation between $D_n4000$ and EW$_\mathrm{[OII]}$}\label{ssection:d4000_oii}

\begin{figure}
\centering
\includegraphics[width=\linewidth]{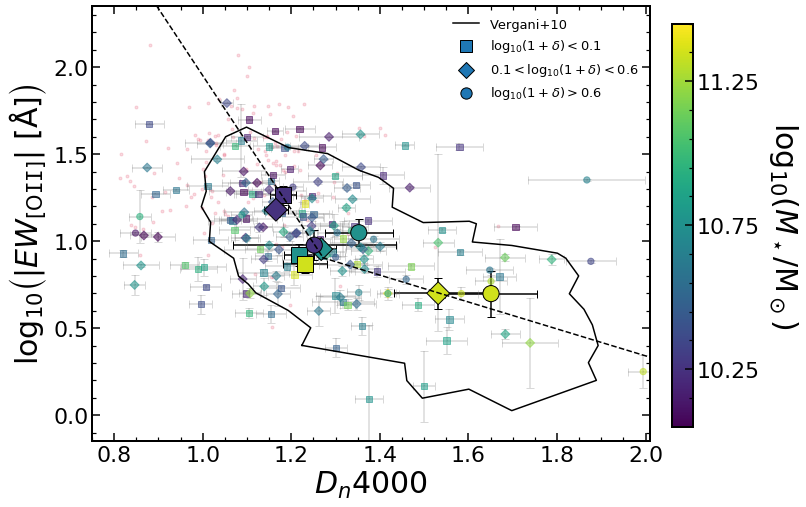}
\caption{The observed relation between $D_n4000$ and $\log_{10}(-EW_\mathrm{[OII]})$ for individual galaxies (small symbols) and median per stellar mass bin (large symbols), colour coded by their stellar mass. \add{The small purple dots show the values for galaxies with $\log_{10}\left(M_\star/\mathrm{M_\odot}\right)<10$.} Different symbols correspond to different over-densities. The dashed lines are linear fits to the data of individual galaxies in two stellar mass subsamples (see text for details): the steeper slope is the fit for galaxies with $10<\log_{10}\left(M_\star/\mathrm{M_\odot}\right)<11$; the shallower slope is the fit for galaxies with $\log_{10}\left(M_\star/\mathrm{M_\odot}\right)>11$. This highlights the underlying anti-correlation between the observed strength of the [O{\sc ii}] emission and the strength of the 4000\AA\ break. We show as a black contour the location of 85\% of the zCOSMOS sample at $0.48<z<1.2$ and with stellar masses greater than $10^{10}\mathrm{M_\odot}$ \citep{vergani2010}.}
\label{fig:D4000_OII_Mass}
\end{figure}

We show in Fig. \ref{fig:D4000_OII_Mass} an anti-correlation between the strength of the 4000 \AA\ break (traced by $D_n4000$) and the [O{\sc ii}] line equivalent width which broadly traces the sSFR \citep{darvish2015}. The observed trend seems to be partially induced by a variation in stellar mass and we find that the most massive galaxies ($\log_{10}\left(M_\star/\mathrm{M_\odot}\right)>11$) in intermediate and high-density regions show an increase in $D_n4000$ while having similar [O{\sc ii}] line equivalent widths as lower density regions. We attempt at qualifying the correlation by fitting two linear models on two different stellar mass regimes

\begin{equation}
Y = \begin{cases}
    (-3.9\pm0.7) \times X + (5.8\pm0.9)  &  \text{if } 10<M<11\\
    (-0.8\pm0.2) \times X+ (1.9\pm0.3) & \text{if } M>11
\end{cases}
\label{eq:oii_d4000}
\end{equation}

\noindent
with $Y=\log_{10}(-\mathrm{EW_{[OII]}})$, $X=D_n4000$, and $M=\log_{10}\left(M_\star/\mathrm{M_\odot}\right)$. This means a steeper slope for the less massive galaxies when compared to the one for the most massive galaxies. We find that the departure from the lower stellar mass relation happens at lower stellar masses for higher density regions, with galaxies in field-like regions never departing from that relation.

This relation is qualitatively similar to what is reported for galaxies more massive than $10^{10}\mathrm{M_\odot}$ in zCOSMOS at $0.48<z<1.2$ by \citet{vergani2010}, although they only use this relation to compare the selection of star-forming, quiescent, and post-starburst galaxies. We interpret this relation as a combination of two phenomena. At lower stellar masses ($<10^{11}\mathrm{M_\odot}$) it is likely a consequence of a declining sSFR with the stellar mass that drives the decrease of [O{\sc ii}] equivalent width and an increase of $D_n4000$. At higher stellar masses there are possible other ionizing mechanisms than star formation alone. We note that the fraction of quiescent galaxies is higher at high stellar masses and also higher in the high-density regions we probe \citeAPA. It is also found in the literature that for such red systems, the [OII] line is likely dominated by AGN/LINER powered emission \citep[e.g.][]{yan2006,lemaux2010}. More recently, resolved studies of galaxies have also revealed that LINER-like emission can be widespread and is thought to be powered by an extended population of post-Asymptotic Red Giant branch stars \citep[see e.g.][]{singh2013, gomes2016, belfiore2016, belfiore2017}. It is also possible that an increased stellar metallicity would result in an increase of $D_n4000$ at fixed  \EWoii, but there is little evidence that metallicity strongly depends on environment \citep[e.g. for gas-phase metallicity][and for stellar metallicity \citealt{harrison2011}]{ellison2009, cooper2008b, darvish2015, sobral2015b, sobral2016, wu2017}. Another likely explanation is that there is an underlying older stellar population in higher density environments (as hinted from Section \ref{ssection:d4000}) with residual star formation producing the observed [O {\sc ii}] emission. This could be seen through a similar break in the sSFR-$D_n4000$ relation, which we do not observe as prominently in our sample. However, we cannot test on the possible LINER-like emission nature of [O{\sc ii}] without spectroscopic observations at longer wavelengths.

\subsection{Star formation activity in different environments}\label{ssection:quiescentAndSF}

\begin{figure}
\centering
\includegraphics[width=\linewidth]{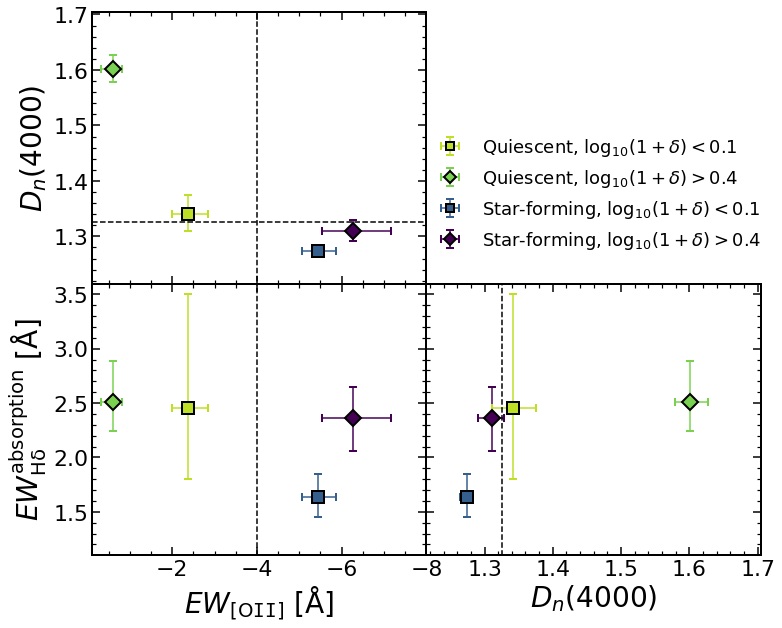}
\caption{We show the properties of composite spectra of quiescent and star-forming galaxies selected through their sSFR (separation at $\log_{10}(sSFR)=-11$, see also \citeprep{PA18}) in low- (diamonds) and high-density regions (squares). The dotted lines are for $D_n4000=1.325$ and EW$_\mathrm{[OII]}=4$ \AA\, and shown to guide the eye. We see that these two populations are separated in $D_n4000$ and EW$_\mathrm{[OII]}$ and that we can also find some environmental effects within each population of galaxies. We find that it is in high-density regions that the quiescent and star-forming populations are more clearly separated.}
\label{fig:stacksProps_QuiescentStarForming}
\end{figure}

Finally, we have also studied the composite spectra of quiescent and star-forming galaxies (more massive than \msun{10} separated at SED-derived $\log_{10}(sSFR)=-11$, see \citeAPA) in low ($\log_{10}(1+\delta)<0.1$) and high ($\log_{10}(1+\delta)>0.4$) density environments and summarize our results in Fig. \ref{fig:stacksProps_QuiescentStarForming} (see also Table \ref{tab:results}). We find that the two populations are separated with $D_n4000$ and EW$_\mathrm{[OII]}$, as expected. Overall, we also find that the difference is larger in high-density regions.

When focusing on H$\delta$ absorption strength, we see stronger absorption in star-forming galaxies in high-density regions than at lower densities. These results are indicative of star-forming galaxies in high-density regions having undergone a recent burst of star formation (or being a mix of normal star-forming galaxies with a post-starburst population), or having dustier star-forming regions \citep[e.g.][]{smail1999}. That scenario would explain similar observed [O{\sc ii}] equivalent widths but a $\sim50\pm20$\% increase in the H$\delta$ absorption strength and only a small $\sim3\pm2$\% increase in $D_n4000$ \citep[see e.g.][]{balogh1999,poggianti1999,mansheim2017}. \add{For field star-forming galaxies we find lower equivalent widths for the H$\delta$ absorption. This may hint at a less bursty star-formation on these systems.} 

From estimates using a single stellar population model with a stellar mass-based stellar metallicity (see Section \ref{ssection:stellar_age}) we find that quenched galaxies in the densest regions are much older (assuming a SSP, $\sim2.2_{-0.4}^{+0.1}$ Gyr) than those in the field ($\sim0.54_{-0.07}^{+0.09}$ Gyr). These differences among quiescent galaxies with the environment are not found by \citet{muzzin2012} or \citet{mosleh2018}. We note that both quiescent samples in low- and high-density regions have similar stellar mass distributions, so we should not expect this to be a simple consequence of an underlying mass-metallicity relation. We discuss further interpretations of our results in Section \ref{ssection:oldpop}.


\section{Discussion}\label{section:discussion}

Our results highlight that both stellar mass and environment play a role in the star formation history of galaxies. \citep[as also reported by e.g.][]{iovino2010, cucciati2010b, peng2010b, li2011, davidzon2016, darvish2016, kawinwanichakij2017}. We see in Fig. \ref{fig:stacksProps_dualMass} that stellar-mass influences the strength of the [O{\sc ii}] in all environments, showing weaker emission for the most massive galaxies, with stellar mass being the main driver of the observed changes. However, we see that the difference between populations of different stellar masses is affected by the environment. A similar result is also seen in H$\delta$ absorption, with field-like regions being the place where most differences are found. This is also seen clearly in $D_n4000$ where the difference among galaxies with different stellar masses depends on the local density, with galaxies in the field being the most similar.

\subsection{Environmental effects on star formation}\label{ssection:filamentEnhance}

Figure \ref{fig:stacksProps_dualMass} shows that for galaxies with stellar masses $10.5<\log_{10}\left(M_\star/\mathrm{M_\odot}\right)<11$ there is a small increase (at $\sim1.7\sigma$ level) in the observed emission strength of [O{\sc ii}] at filament-like densities when compared to field and cluster regions (see also Fig. \ref{fig:stack_properties} for stacks with all galaxies more massive than $10^{10}\mathrm{M_\odot}$). Despite this trend hinting to a slightly higher sSFR at intermediate densities, it is not evidenced enough of any environmental impact on SFR at these stellar masses.

Considering the sample at $10.0<\log_{10}\left(M_\star/\mathrm{M_\odot}\right)<10.5$ we find an overall decrease in the \EWoii with increasing density, which is also seen for similar stellar masses in \citet[][]{tomczak2018}. On the high stellar mass end of our sample, we see no effect of the environment on the observed \EWoii. This overall effect on the median sSFR can be linked to the results on the quenched fraction ($f_Q$) shown in \citeAPA, where we find that at high stellar masses ($>$\msun{11}) there is no dependence effect on $f_Q$ while at lower stellar masses there is an increase from intermediate- to high-density regions \citep[see also e.g.][]{lemaux2018}. Naturally, an increase of the quenched fraction \add{(from $\sim$10\% to $\sim$50\% of the sample from low and intermediate to high-density regions, see Table \ref{tab:results})} results in a lower median sSFR value for the population. Both these results point to an environmental dependence of star formation at $10.0<\log_{10}\left(M_\star/\mathrm{M_\odot}\right)<10.5$ \citep[see also e.g.][]{peng2010}, while for more massive galaxies we see no major effect. This would support a scenario in which mass quenching mechanisms act to suppress star formation at high stellar masses, and local environment has a negligible impact on star formation \citep[e.g.][]{peng2010,peng2012}, but more recent studies see a different picture for massive galaxies in dense environments \citep[e.g.][]{darvish2016}. 

A scenario that can explain these results must account for enhancement of star formation activity at intermediate densities and then some quenching mechanism that acts as galaxies move towards higher densities (which may be connected). This can be thought of as filaments being regions of higher probabilities for gas-rich galaxies to interact \citep[e.g.][]{moss2006,perez2009,li2009,tonnesen2012,darvish2014,malavasi2017} promoting compression of gas clouds and a peak in the SFR of the galaxy \citep[e.g.][]{mihos1996,kewley2006,ellison2008, gallazzi2009,bekki2009,owers2012,roediger2014}. Since we are looking at a superstructure composed of several sub-clusters, it is also possible that activity related to cluster-cluster interactions are capable of enhancing star formation as well \citep[e.g.][]{stroe2014,stroe2015a,sobral2015b}, although this might not always be the case \citep[see e.g.][]{mansheim2017b}.

\subsection{Older population prevalence influenced by the environment}\label{ssection:oldpop}

While we find in Fig. \ref{fig:stack_properties} that $D_n4000$ strongly correlates with stellar mass, we show in Fig. \ref{fig:stacksProps_dualMass} that the strength of such correlation is strongly dependent on the environment. The fact that cluster galaxies have on average stronger 4000 \AA\ breaks were already reported in other studies \citep[see e.g.][]{muzzin2012}, but we extend to a range of densities that is complementary to the sample of \citet[][]{muzzin2012}, targeting rich clusters and the galaxies around them.

To explain the observed trend, we need that the time passed since the last relevant episode of star formation be different at each environment \citep[see also e.g.][]{rettura2010,rettura2011,darvish2016}. When looking at the most massive galaxies ($\log_{10}\left(M_\star/\mathrm{M_\odot}\right)>11$) we find that they have older stellar populations in high-density regions when compared to the lower density region counterparts ($5_{-2}^{+3}$ Gyr compared to $0.64_{-0.07}^{+0.06}$ Gyr at lower densities\footnotemark{}\footnotetext{We note that this age estimate is based on an SSP with a median stellar metallicity derived for individual subsamples, see Section \ref{ssection:stellar_age}.}). {\citeAPA\ showed} that, at these stellar masses, the quenched fraction is not changing with the environment, so it is not simply a larger fraction of red galaxies that can reproduce the age differences. This would then require that massive galaxies in the high-density regions have become passive earlier than their field counterparts, which can be a consequence of denser regions having collapsed first and thus galaxies in such regions formed at earlier times \citep[e.g.][]{thomas2005,nelan2005}. It is also consistent with the observed age difference for quenched galaxies in the field and cluster-like densities ($\sim0.54_{-0.07}^{+0.09}$  Gyr and $\sim2.2_{-0.4}^{+0.1}$ Gyr, respectively). 

The lack of recent star formation in higher density regions can be linked to the gas accretion being lower in dense regions \citep[e.g.][]{vandevoort2017}. Due to the prevalence of gas stripping in dense regions \citep[e.g.][]{bahe2013,jaffee2015,poggianti2016}, also seen in the lack of gas-rich galaxies in clusters  \citep[e.g.][]{boselli2016}, there is little fuel to new star-forming episodes for massive galaxies. In lower-density environments (fields, filaments, or groups) it is more likely that gas can funnel to the massive galaxies from minor mergers or tidal interactions with gas-rich satellites \citep[][]{poggianti2016}. This is also consistent with the pre-processing of galaxies in filaments and groups (gas loss and quiescence) before in-fall in the cluster regions \citep[][]{wetzel2013,haines2015}. The ram pressure from the denser regions can also act as an enhancement of star formation prior to quenching \citep[see Section \ref{ssection:filamentEnhance} and e.g.][]{poggianti2016}. Thus, massive galaxies in high-density regions should grow through mergers of gas-poor galaxies \citep[dry mergers, see e.g.][]{khochfar2003,khochfar2009, mcintosh2008, lin2010, davidzon2016, martin2017} in order to build up their mass while maintaining an older stellar population. In this scenario, the dependence of the $D_n4000$-stellar mass on the environment is explained by the fraction of gas-rich galaxies that are available for merging at each density, being less common at higher densities \citep[e.g.][]{boselli2016}. This is also in agreement to numerous reported trends of the quenched fraction (equivalent to the number of gas-poor galaxies) with environment \citep[][\!; \citeAPA]{peng2010b, cucciati2010b, sobral2011, muzzin2012, darvish2016}.

As we move to stellar masses $10<\log_{10}\left(M_\star/\mathrm{M_\odot}\right)<11$, the impact of environment is less pronounced (differences in $D_n4000$ between low- and high-density environments is smaller), although we still see slightly older stellar populations in higher density environments ($\sim$0.6 Gyr older than in low-density environments). In this case, we would require either that galaxies are still forming new stars or that if they are quenched, it happened more recently. Since we observe a rise of the quenched fraction with local density for these stellar masses in \citeAPA, we would favour the latter scenario. 

We can reconcile the environmental effects seen at high stellar masses in $D_n4000$ with those seen only at lower stellar masses in \EWoii and $f_Q$ discussed in Section \ref{ssection:filamentEnhance}. The values of \EWoii and $f_Q$ (directly related to a galaxy' sSFR by our definition) are probing star formation timescales of 10-100 Myr \citep[e.g.][]{couch1987, salim2007, cunha2008},  while the value for $D_n4000$ can probe the lack of recent star formation on longer timescales (typically a few Gyr, see Section \ref{ssection:stellar_age}). These different tracers can then be thought of a difference between a galaxy being quenched or not (\emph{instantaneous} star formation traced by [OII]) and the time that has passed since that quenching happened (traced by $D_n4000$). A simple explanation is possibly linked to the formation epoch of massive galaxies in different environments \citep[e.g.][]{thomas2005,nelan2005}. Galaxies in cluster cores are formed earlier (due to earlier collapse of over-dense regions) and therefore become passive earlier (having low sSFR and high $D_n4000$ due to the median stellar age passively evolving because of the absence of further episodes of star formation) while galaxies in lower density regions can be recently quenched (having low sSFR and relatively low $D_n4000$)  since their regions have collapsed later in cosmic time. It is also possible that faster quenching in high-density regions \citep[e.g.][]{foltz2018,socolovsky2018} could mimic the observations if galaxies formed at similar cosmic times. These scenarios would explain the higher values of $D_n4000$ and lower values of H$\delta$ absorption equivalent width for high-density quiescent galaxies when compared to the lower density counterparts. Both scenarios require that higher density environments ultimately suppress or do not allow for new star formation episodes in galaxies (which might happen through bursts or in a continuous decline).

We note that a higher metallicity in high-density regions \citep[not encoded in our assumed stellar mass-metallicity relation in Section \ref{ssection:stellar_age}, e.g.][]{sobral2015b} could mitigate some of the observed differences in our age estimates. If we consider a metallicity of 2.5$Z_\odot$ (4.5 times higher than the expected value from the median stellar mass at this redshift), we would have an estimate of age$\sim1.7$ Gyr for quiescent galaxies in high-density regions. However, other studies find only marginal differences in gas-phase metallicity between field and cluster quiescent galaxies \citep[e.g.][see also e.g. \citealt{cooper2008b,darvish2015,sobral2016,wu2017} for similar results on star-forming galaxies]{ellison2009}, also seen in stellar metallicity \citep[e.g.][]{harrison2011}, making it unlikely that a metallicity dependence on environment can explain the observed differences.


\section{Conclusions}\label{section:conclusions}

We have presented the spectroscopic properties of 466 galaxies in and around a $z\sim0.84$ superstructure in the COSMOS field targeted with the VIS$^3$COS survey \citep[][]{paulino-afonso2018a}. We explore the spectral properties of galaxies and relate those to their stellar mass and environment, by measuring and interpreting [O{\sc ii}], H$\delta$, and $D_n4000$. We use [O{\sc ii}] equivalent width as a tracer of sSFR, H$\delta$ as a tracer of current episodes (from emission) or recent bursts (from absorption) of star formation, and $D_n4000$ as a tracer of the average age of the stellar population. We present results both on individual galaxies and on composite spectra to evaluate the relative importance of stellar mass and/or the environment in the build-up of stellar populations in galaxies. Our main results are:

\begin{itemize}
\setlength\itemsep{0.5em}
\item \add{We find no significant dependence of H$\delta$ absorption or emission components on the environment. We find that both H$\delta$ absorption or emission decrease with increasing stellar mass.}
\item The [O{\sc ii}]$\lambda$3727 absolute line equivalent width decreases \add{by a factor of $\sim$3 from $\sim$\msun{10.25} to $\sim$\msun{11.25}}. We observe this decrease in all environments which trend is mostly a consequence of the underlying main sequence of star-forming galaxies.
\item We find $D_n4000$ to increase with an increasing stellar mass in all environments. For stellar masses $10.0<\log_{10}\left(M_\star/\mathrm{M_\odot}\right)<10.5$ we see an impact of environment on the average stellar age ($0.32_{-0.05}^{+0.06}$ Gyr to $1.1\pm0.2$ Gyr, from low- to high-density regions). For the most massive galaxies ($\log_{10}\left(M_\star/\mathrm{M_\odot}\right)>11$) the difference is much larger ($0.64_{-0.07}^{+0.06}$ Gyr to $5_{-2}^{+3}$ Gyr\footnotemark{}\footnotetext{We note that this age estimate is based on a single SSP. See Section \ref{ssection:oldpop}.}).
\item There is an anti-correlation between $\log_{10}\left(-EW_\mathrm{[OII]}\right)$ and $D_n4000$ \citep[also seen in e.g.][]{vergani2010} which is mostly a consequence of the underlying correlations of these quantities with stellar mass. We find that the most massive galaxies ($\log_{10}\left(M_\star/\mathrm{M_\odot}\right)>11$) in intermediate and high-density regions have higher $D_n4000$ while showing similar EW\textsubscript{[O{\sc ii}]} to lower density regions. This may hint at a different ionizing mechanism in high stellar mass galaxies operating in denser environments or be a consequence of older stellar populations residing in such regions.
\item We find an increase in the [O{\sc ii}] equivalent width at intermediate densities for intermediate stellar mass galaxies ($10.5<\log_{10}\left(M_\star/\mathrm{M_\odot}\right)<11$) which may point to episodes of enhanced star formation (more stars formed per stellar mass) on timescales around 10 Myr.
\end{itemize}

Based on our results on $D_n4000$, we hypothesize that the most massive galaxies ($\log_{10}\left(M_\star/\mathrm{M_\odot}\right)>11$) have ceased star formation earlier (by a few Gyr, depending on the stellar metallicity and assumed star formation history) in high-density environments than their field counterparts. Lower stellar mass galaxies ($10<\log_{10}\left(M_\star/\mathrm{M_\odot}\right)<11$) need to have quenched more recently at similar environments (or still have ongoing lower levels of star formation), since they have signs of younger stellar populations when compared to the most massive sample. The observed older stellar populations of massive galaxies in high-density environments point to a lack of recent episodes of significant star formation. This is compatible with a scenario where either all stars formed in-\textit{situ} and earlier or that they likely growth mechanism through dry merging events. In lower-density environments, they are either continuously forming new stars (at lower rates) or experiencing merging events with gas-rich galaxies fuelling new episodes of star formation. Such a scenario is required to explain the dependence of $D_n4000$ on the environment at these stellar masses.


\begin{acknowledgements}
We thank the anonymous referee for the insightful and useful comments that helped improve the quality and readability of the paper. This work was supported by Funda\c{c}\~{a}o para a Ci\^{e}ncia e a Tecnologia (FCT) through the research grant UID/FIS/04434/2013. APA, PhD::SPACE fellow, acknowledges support from the FCT through the fellowship PD/BD/52706/2014. DS acknowledges financial support from Lancaster University through an Early Career Internal Grant A100679. BD acknowledges financial support from NASA through the Astrophysics Data Analysis Program (ADAP), grant number NNX12AE20G, and the National Science Foundation, grant number 1716907. IRS acknowledges support from the ERC Advanced Grant DUSTYGAL (321334), STFC (ST/P000541/1) and a Royal Society/Wolfson Merit award. PNB is grateful for support from STFC (ST/M001229/1).

This work was only possible by the use of the following \textsc{python} packages: NumPy \& SciPy \citep{walt2011,jones2001}, Matplotlib \citep{hunter2007}, Astropy \citep{robitaille2013}, and EzGal \citep{mancone2012}.
\end{acknowledgements}

\bibliographystyle{aa}
\bibliography{refs}

\begin{appendix}

\section{Detailed spectral stack results}\label{app:stack_results}

In this section, we briefly describe some of the observed features in the composite spectra and refer any quantitative analysis to Section \ref{ssection:linemeasures}. 

We show in Fig. \ref{fig:stack_mass_full} the composite spectra in bins of stellar mass. We observe a strong decrease in [O{\sc ii}] line flux from low to high stellar masses. We also see the relative strength of the two doublet lines changing with stellar mass (a quantitative analysis on the electron density estimates will be subject of a future paper). We find a decrease of the emission strength of Balmer lines (H$\gamma$, H$\delta$, H$\epsilon$, H$\zeta$, H$\theta$) with increasing stellar mass. At the same time, the prominence of the absorption features is increasingly noticeable at higher stellar masses. We also note the presence of [Ne{\sc iii}] emission in some spectra which will be the subject of a forthcoming paper based on the VIS$^{3}$COS survey.

In Fig.\ref{fig:stack_density_full} we show our findings of the stacked spectra in bins of over-density. In terms of the [O{\sc ii}] emission, we find a decreasing line strength from low to high-density regions (see also \citeAPA). In terms of the H$\delta$ line, we also see a dependence on local density. We find an increase in the absorption strength from low- to high-density and a decrease of the emission component.

Finally, in Fig. \ref{fig:stack_sfr_full} we show the composite spectra binned by SFR. As expected, the [O{\sc ii}] emission is stronger for high SFR galaxies. In terms of their H$\delta$ absorption, we see a stronger absorption in higher SFR galaxies.

\begin{figure*}
\centering
\includegraphics[width=0.95\textwidth]{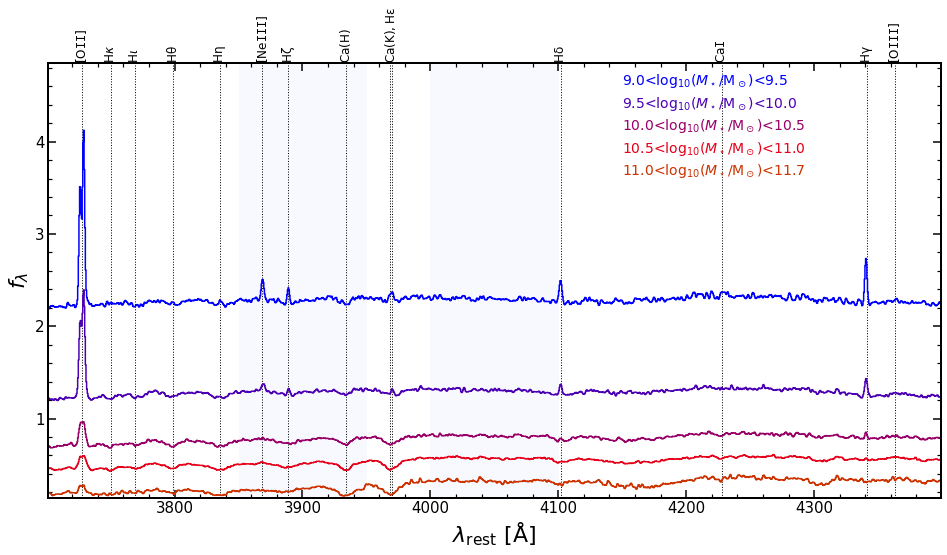}
\caption{Resulting median composite spectra normalized at 4150-4300 \AA\ and associated error (solid line + shaded region) in bins of stellar mass (low to high stellar mass from top to bottom). We apply a vertical offset for visualization purposes. We highlight with vertical lines the strongest features that we see on our spectra. The light grey vertical stripes show the spectral ranges which are used to compute $D_n4000$.}
\label{fig:stack_mass_full}
\end{figure*}

\begin{figure*}
\centering
\includegraphics[width=0.95\textwidth]{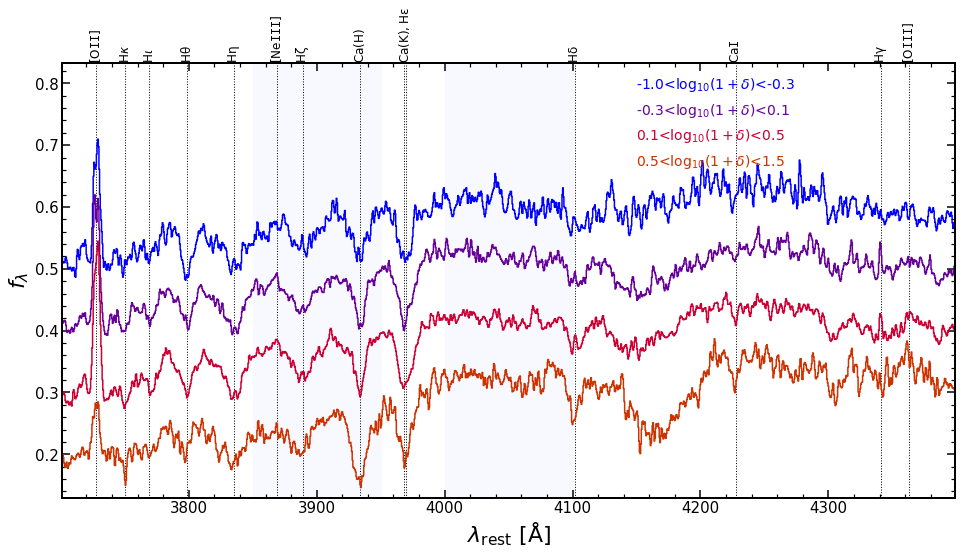}
\caption{Resulting median composite spectra stacks normalized at 4150-4300 \AA\ and associated error (solid line + shaded region) in bins of local over-density (low- to high-density regions from top to bottom) for galaxies with stellar masses greater than $10^{10}\mathrm{M_\odot}$. We apply a vertical offset for visualization purposes. We highlight with vertical lines the strongest features that we see on our spectra. The light grey vertical stripes show the spectral ranges which are used to compute $D_n4000$.}
\label{fig:stack_density_full}
\end{figure*}

\begin{figure*}
\centering
\includegraphics[width=0.95\textwidth]{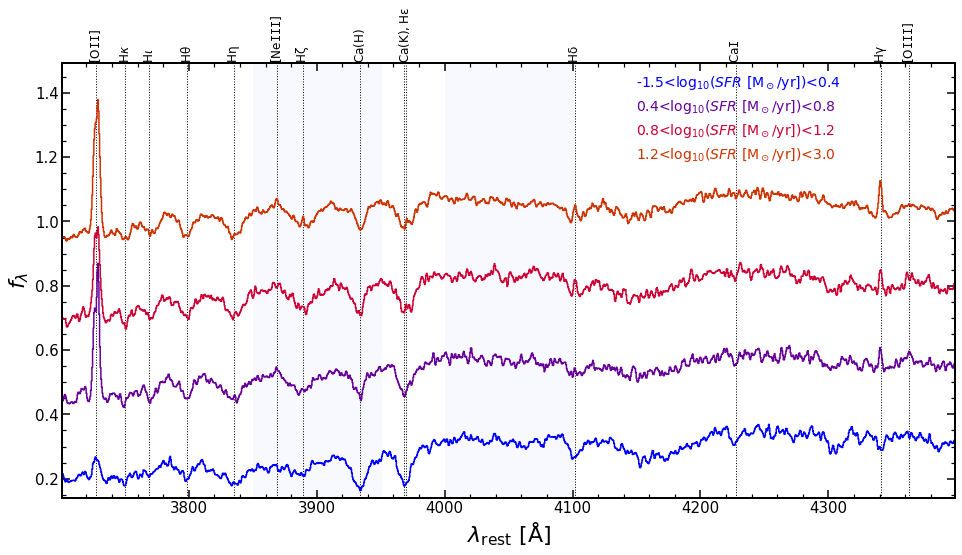}
\caption{Resulting median composite spectra normalized at 4150-4300 \AA\  and associated error (solid line + shaded region) in bins of star formation rate (high to low SFR from top to bottom) for galaxies with stellar masses greater than $10^{10}\mathrm{M_\odot}$. We apply a vertical offset for visualization purposes. We highlight with vertical lines the strongest features that we see on our spectra. The light grey vertical stripes show the spectral ranges which are used to compute $D_n4000$.}
\label{fig:stack_sfr_full}
\end{figure*}

\begin{table*}
\centering
\caption{Summary table with the results for all spectroscopic indices in composite spectra shown in this paper. If only one variable is defined, it means that there is only a single selection for that row. Samples not selected in stellar mass in variables 1 and/or 2 have a minimum stellar-mass limit of $10^{10}\mathrm{M_\odot}$. \add{The third column shows the number of galaxies for each selection and the fourth column, the number of star-forming galaxies ($\log_{10}(sSFR)>-11$)}. The last column shows the stellar age for a single stellar population from \citet{bruzual2003} \add{derived as explained in Section \ref{ssection:stellar_age}}. We opt not to show errors for the derived ages since the systematics due to our choice of metallicity and star formation history are much larger than those derived from observational errors on $D_n4000$.}
\label{tab:results}
\resizebox{\textwidth}{!}{
\begin{tabular}{ccccccccc}
Variable 1 & Variable 2 & $N_\mathrm{all}$ & $N_\mathrm{SF}$ & [O{\sc ii}] & H$\delta_\mathrm{emi}$ & H$\delta_\mathrm{abs}$ & $D_n4000$ & $t_\mathrm{SSP}$ \\
 & & & & \AA & \AA & \AA & & Gyr \\
\hline
$9.0 \leq \log_{10}(M/\mathrm{M_\odot}) < 9.5$ & - & 49 & 49  &  $-27\pm2$  &  $-2.5\pm0.3$  &  N/A  &  $1.09\pm0.01$  &  $0.04_{-0.01}^{+0.03}$\\
$9.5 \leq \log_{10}(M/\mathrm{M_\odot}) < 10.0$ & - & 111 & 108  &  $-22\pm2$  &  $-1.01_{-0.10}^{+0.09}$  &  $3.8_{-0.7}^{+0.6}$  &  $1.117\pm0.009$  &  $0.06\pm0.02$\\
$10.0 \leq \log_{10}(M/\mathrm{M_\odot}) < 10.5$ & - & 107 & 95  &  $-6.7_{-0.8}^{+0.7}$  &  $-0.6_{-0.2}^{+0.3}$  &  $2.7\pm0.2$  &  $1.27_{-0.02}^{+0.01}$  &  $0.42_{-0.04}^{+0.03}$\\
$10.5 \leq \log_{10}(M/\mathrm{M_\odot}) < 11.0$ & - & 148 & 121  &  $-4.3\pm0.3$  &  $-0.39_{-0.08}^{+0.07}$  &  $1.9\pm0.2$  &  $1.365\pm0.010$  &  $0.61\pm0.03$\\
$11.0 \leq \log_{10}(M/\mathrm{M_\odot}) < 11.7$ & - & 40 & 19  &  $-2.0_{-0.4}^{+0.3}$  &  $-0.1\pm0.1$  &  $1.1_{-0.2}^{+0.3}$  &  $1.56\pm0.03$  &  $1.3_{-0.2}^{+0.4}$\\

\hline
$-1.0 \leq \log_{10}(1+\delta) < -0.3$ & - & 42 & 32  &  $-4.0_{-0.5}^{+0.6}$  &  $-0.4\pm0.2$  &  $2.3_{-0.5}^{+0.7}$  &  $1.26\pm0.02$  &  $0.35\pm0.04$\\
$-0.3 \leq \log_{10}(1+\delta) < 0.1$ & - & 91 & 78  &  $-5.3_{-0.6}^{+0.5}$  &  $-0.3_{-0.2}^{+0.1}$  &  $1.6\pm0.2$  &  $1.31\pm0.01$  &  $0.46\pm0.03$\\
$0.1 \leq \log_{10}(1+\delta) < 0.5$ & - & 101 & 89  &  $-6.5_{-0.4}^{+0.5}$  &  $-0.5_{-0.2}^{+0.1}$  &  $2.1\pm0.2$  &  $1.31\pm0.01$  &  $0.47\pm0.03$\\
$0.5 \leq \log_{10}(1+\delta) < 1.5$ & - & 61 & 36  &  $-2.7_{-0.5}^{+0.3}$  &  $0.0_{-0.4}^{+0.1}$  &  $2.2_{-0.3}^{+0.4}$  &  $1.51\pm0.02$  &  $1.1\pm0.1$\\

\hline
 Field & - & 121 & 101  &  $-4.6_{-0.5}^{+0.4}$  &  $-0.48_{-0.09}^{+0.08}$  &  $2.3\pm0.2$  &  $1.31\pm0.01$  &  $0.48\pm0.03$\\
 Filament & - & 136 & 106  &  $-4.4_{-0.4}^{+0.3}$  &  $-0.5_{-0.2}^{+0.1}$  &  $1.9\pm0.2$  &  $1.34\pm0.01$  &  $0.56_{-0.03}^{+0.04}$\\
 Cluster & - & 38 & 28  &  $-3.4\pm0.5$  &  $0.0_{-0.2}^{+0.1}$  &  $1.9_{-0.3}^{+0.4}$  &  $1.52\pm0.03$  &  $1.1_{-0.1}^{+0.2}$\\

\hline
$-1.5 \leq \log_{10}(SFR\ \mathrm{M_\odot yr^{-1}}) < 0.4$ & - & 98 & 38  &  $-1.4\pm0.2$  &  $-0.1_{-0.3}^{+0.1}$  &  $1.9_{-0.3}^{+0.2}$  &  $1.47\pm0.01$  &  $0.92_{-0.07}^{+0.08}$\\
$0.4 \leq \log_{10}(SFR\ \mathrm{M_\odot yr^{-1}}) < 0.8$ & - & 48 & 48  &  $-7.9_{-0.7}^{+0.9}$  &  $-0.19_{-0.07}^{+0.06}$  &  $1.4\pm0.2$  &  $1.32\pm0.02$  &  $0.50\pm0.04$\\
$0.8 \leq \log_{10}(SFR\ \mathrm{M_\odot yr^{-1}}) < 1.2$ & - & 48 & 48  &  $-6.9_{-0.7}^{+0.4}$  &  $-0.7_{-0.2}^{+0.1}$  &  $2.4\pm0.2$  &  $1.29_{-0.02}^{+0.01}$  &  $0.44_{-0.04}^{+0.03}$\\
$1.2 \leq \log_{10}(SFR\ \mathrm{M_\odot yr^{-1}}) < 3.0$ & - & 101 & 101  &  $-10.3\pm0.5$  &  $-0.8\pm0.1$  &  $2.6_{-0.3}^{+0.2}$  &  $1.19\pm0.01$  &  $0.22\pm0.02$\\

\hline
$-1.0 \leq \log_{10}(1+\delta) < 0.1$  & $10.0 \leq \log_{10}(M/\mathrm{M_\odot}) < 10.5$ & 49 & 44  &  $-9_{-2}^{+1}$  &  $-0.6_{-0.1}^{+0.2}$  &  $2.6\pm0.4$  &  $1.23_{-0.02}^{+0.01}$  &  $0.32_{-0.05}^{+0.06}$\\
$-1.0 \leq \log_{10}(1+\delta) < 0.1$  & $10.5 \leq \log_{10}(M/\mathrm{M_\odot}) < 11.0$ & 63 & 55  &  $-3.7_{-0.5}^{+0.4}$  &  $-0.5\pm0.2$  &  $1.7\pm0.2$  &  $1.31_{-0.02}^{+0.01}$  &  $0.46\pm0.03$\\
$-1.0 \leq \log_{10}(1+\delta) < 0.1$  & $11.0 \leq \log_{10}(M/\mathrm{M_\odot}) < 12.0$ & 21 & 11  &  $-3.4_{-0.4}^{+0.5}$  &  $0.00_{-0.04}^{+0.01}$  &  $0.7_{-0.7}^{+0.5}$  &  $1.39_{-0.02}^{+0.03}$  &  $0.64_{-0.07}^{+0.06}$\\
$0.1 \leq \log_{10}(1+\delta) < 0.6$  & $10.0 \leq \log_{10}(M/\mathrm{M_\odot}) < 10.5$ & 50 & 47  &  $-8.0_{-0.9}^{+0.7}$  &  $-0.4_{-0.3}^{+0.1}$  &  $1.9\pm0.3$  &  $1.23\pm0.02$  &  $0.30_{-0.04}^{+0.05}$\\
$0.1 \leq \log_{10}(1+\delta) < 0.6$  & $10.5 \leq \log_{10}(M/\mathrm{M_\odot}) < 11.0$ & 57 & 48  &  $-5.2\pm0.4$  &  $-0.4_{-0.4}^{+0.1}$  &  $1.7_{-0.3}^{+0.4}$  &  $1.36\pm0.02$  &  $0.61_{-0.05}^{+0.04}$\\
$0.1 \leq \log_{10}(1+\delta) < 0.6$  & $11.0 \leq \log_{10}(M/\mathrm{M_\odot}) < 12.0$ & 12 & 4  &  $-2_{-2}^{+1}$  &  $-0.2_{-0.1}^{+0.2}$  &  $1.8_{-0.2}^{+0.3}$  &  $1.54_{-0.06}^{+0.05}$  &  $1.2_{-0.3}^{+0.7}$\\
$0.6 \leq \log_{10}(1+\delta) < 2.0$  & $10.0 \leq \log_{10}(M/\mathrm{M_\odot}) < 10.5$ & 8 & 4  &  $-0_{-4}^{+1}$  &  $-0.6_{-0.1}^{+0.6}$  &  $2_{-2}^{+1}$  &  $1.47\pm0.05$  &  $1.1\pm0.2$\\
$0.6 \leq \log_{10}(1+\delta) < 2.0$  & $10.5 \leq \log_{10}(M/\mathrm{M_\odot}) < 11.0$ & 28 & 18  &  $-4.0_{-0.6}^{+0.5}$  &  N/A  &  $1.7_{-0.3}^{+0.4}$  &  $1.48\pm0.02$  &  $0.9\pm0.1$\\
$0.6 \leq \log_{10}(1+\delta) < 2.0$  & $11.0 \leq \log_{10}(M/\mathrm{M_\odot}) < 12.0$ & 7 & 4  &  $-2_{-2}^{+1}$  &  $0.0_{-0.4}^{+0.1}$  &  $1\pm1$  &  $1.8\pm0.1$  &  $5_{-2}^{+3}$\\

\hline
$\log_{10}(sSFR)<-11$  & $\log_{10}(1+\delta)<0.1$ & 25 & 0  &  $-2.4_{-0.5}^{+0.4}$  &  $0.0_{-0.5}^{+0.1}$  &  $2\pm1$  &  $1.34\pm0.03$  &  $0.54_{-0.07}^{+0.09}$\\
$\log_{10}(sSFR)<-11$  & $\log_{10}(1+\delta)>0.4$ & 29 & 0  &  $-0.6_{-0.2}^{+0.3}$  &  N/A  &  $2.5_{-0.3}^{+0.4}$  &  $1.60_{-0.02}^{+0.03}$  &  $2.2_{-0.4}^{+0.1}$\\
$\log_{10}(sSFR)>-11$  & $\log_{10}(1+\delta)<0.1$ & 234 & 234  &  $-5.5\pm0.4$  &  $-0.3\pm0.1$  &  $1.6\pm0.2$  &  $1.274\pm0.010$  &  $0.39\pm0.02$\\
$\log_{10}(sSFR)>-11$  & $\log_{10}(1+\delta)>0.4$ & 80 & 80  &  $-6.3_{-0.9}^{+0.7}$  &  $-0.6\pm0.2$  &  $2.4\pm0.3$  &  $1.31\pm0.02$  &  $0.47_{-0.04}^{+0.05}$\\

\hline
Field  & $10.0 \leq \log_{10}(M/\mathrm{M_\odot}) < 10.5$ & 50 & 45  &  $-6_{-2}^{+1}$  &  $-0.9\pm0.2$  &  $3.2\pm0.3$  &  $1.31_{-0.03}^{+0.01}$  &  $0.6\pm0.1$\\
Field  & $10.5 \leq \log_{10}(M/\mathrm{M_\odot}) < 11.0$ & 55 & 50  &  $-3.9\pm0.4$  &  $-0.1\pm0.1$  &  $2.0\pm0.3$  &  $1.30\pm0.01$  &  $0.45\pm0.03$\\
Field  & $11.0 \leq \log_{10}(M/\mathrm{M_\odot}) < 12.0$ & 16 & 6  &  $-3.7_{-0.3}^{+0.4}$  &  N/A  &  N/A  &  $1.35\pm0.03$  &  $0.55_{-0.07}^{+0.08}$\\
Filament  & $10.0 \leq \log_{10}(M/\mathrm{M_\odot}) < 10.5$ & 51 & 46  &  $-8\pm1$  &  $-0.5_{-0.2}^{+0.1}$  &  $1.8_{-0.3}^{+0.4}$  &  $1.20\pm0.02$  &  $0.24\pm0.04$\\
Filament  & $10.5 \leq \log_{10}(M/\mathrm{M_\odot}) < 11.0$ & 66 & 50  &  $-3.8\pm0.3$  &  $-0.5\pm0.2$  &  $2.0\pm0.3$  &  $1.39_{-0.02}^{+0.01}$  &  $0.68\pm0.03$\\
Filament  & $11.0 \leq \log_{10}(M/\mathrm{M_\odot}) < 12.0$ & 19 & 10  &  $-1.6_{-0.5}^{+0.3}$  &  $-0.15_{-0.09}^{+0.05}$  &  $1.7\pm0.4$  &  $1.60\pm0.03$  &  $2.0_{-0.5}^{+0.2}$\\
Cluster  & $10.0 \leq \log_{10}(M/\mathrm{M_\odot}) < 10.5$ & 6 & 4  &  $-4_{-3}^{+1}$  &  $-0.3_{-0.2}^{+0.3}$  &  $4_{-1}^{+6}$  &  $1.5\pm0.1$  &  $1\pm1$\\
Cluster  & $10.5 \leq \log_{10}(M/\mathrm{M_\odot}) < 11.0$ & 27 & 21  &  $-3.0_{-0.5}^{+0.6}$  &  $-0.1_{-0.5}^{+0.1}$  &  $2.0\pm0.6$  &  $1.46\pm0.03$  &  $0.9_{-0.1}^{+0.2}$\\
Cluster  & $11.0 \leq \log_{10}(M/\mathrm{M_\odot}) < 12.0$ & 5 & 3  &  $-1\pm1$  &  N/A  &  $2_{-1}^{+6}$  &  $2.0_{-0.3}^{+0.1}$  &  $9_{-6}^{+1}$\\

\hline
\end{tabular}}
\end{table*}

\end{appendix}
\end{document}